\documentclass[11pt,prd,showpacs,showkeys]{revtex4}

\usepackage{mathrsfs,amsbsy,amssymb,latexsym,amsfonts,amsmath}
\usepackage[dvipdfm]{graphicx,color}
\usepackage{wrapfig}
\usepackage{amsbsy}
\usepackage{alltt}

\makeatletter
	
	\@addtoreset{equation}{section}
\makeatother

\begin{document}

\begin{flushright}
OIQP-12-10
\end{flushright}

\title{A Novel String Field Theory Solving String Theory \\ by Liberating Left and Right Movers}

\author{Holger B. {\sc Nielsen}}
\affiliation{Niels Bohr Institute, University of Copenhagen, \\
17 Belgdamsvej, DK 2100 Denmark\footnote{email: hbech@nbi.dk}}

\author{Masao {\sc Ninomiya}}
\affiliation{Okayama Institute for Quantum Physics, Kyoyama 1-9-1, Okayama 700-0015, Japan.\footnote{email:msninomiya@gmail.com }}

\begin{abstract}

\vspace{0.2cm}

\begin{center}
{\bf abstract}
\end{center} 

\vspace{-0.2cm}

We put forward ideas to a novel string field theory based on making some ``objects'' that essentially describe ``liberated'' left- and right- mover fields $X^{\mu}_{L}(\tau + \sigma)$ and $X^{\mu}_{R}(\tau - \sigma)$ on the string. Our novel string field theory is completely definitely different from any other string theory in as far as a ``null set'' of information in the string field theory Fock space has been removed relatively, to the usual string field theories. So our theory is definitely new. The main progress is that we manage to make our novel string field theory provide the correct mass square spectrum for the string. We finally suggest how to obtain the Veneziano amplitude in our model.
\end{abstract}



\maketitle

\section{Introduction}
\begin{figure}[!htb]
\begin{center}
 \includegraphics[width=0.6\textwidth]{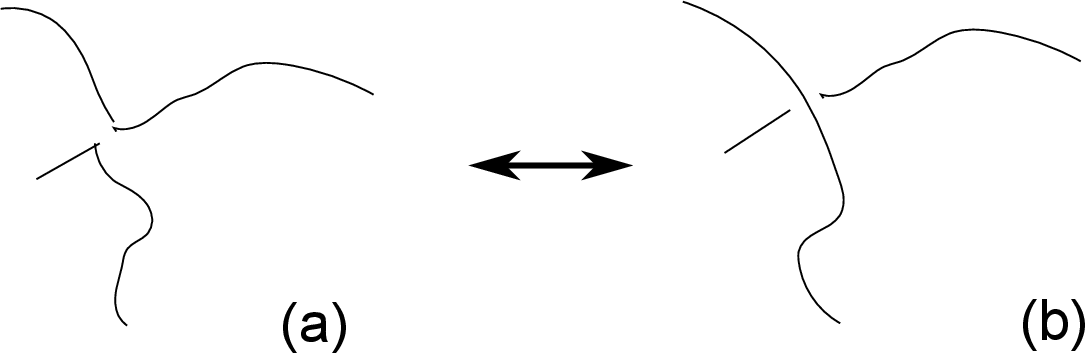}%
 \end{center}
  \caption{In the very moment a scattering takes place for here a couple of open strings except that the couple of strings get split
           and the pieces united in a different way. But wherever there is some string bit it remains on both Figure  \ref{Figure1}(a) and \ref {Figure1}(b). \label{Figure1}}
\end{figure}

In the light of the great possible hopes for superstring theory - which means that we have \underline{many} strings around - the world might have to be described indeed by a string field theory (then with superstrings). One of the great achievements of (super) string theory should be that it is \underline{``finite''}, meaning that the usual ultraviolet divergencies of quantum field theories are avoided. That is so when the dimension is $9+1$ for the superstring theory and $25+1$ for the bosonic string theory \cite{12}-\cite{17}. 
In our early works \cite{1}, \cite{2} we put forward ideas towards novel string field theory (= theory of second quantized strings). Our type of string field theory is guaranteed to be different from the string field theories on the market \cite{3}-\cite{11}. This is so because in our model we do not attach physical significance to how pieces of strings are glued together (see Figure \ref{Figure1}), so that states of several strings which cannot be distinguished from each other by looking everywhere for whether there is one string or another present are considered by \underline{us} the same several string state. Contrary to those of Kaku-Kikkawa and Witten et al. could have different states that everywhere in target space looks the same but which differ by how the string pieces continue into each other.

\subsection{Review of Our Model Starting from Strings}

For the reader which is not familiar with our earlier works \cite{1}, \cite{2} we would 
here explain our idea of a string field theory starting from considering strings:

Since we have, as just mentioned, decided to ignore how different bits 
of strings hang together - and to let it be rather given by the 
suggestion that pieces touching may hang together (it were this 
ignorance of some information present in usual string field theories 
that should guarantee our theory to be different form the 
usual ones) - we shall be satisfied by putting into the string field theory 
we have constructed only local information about how the strings are directed 
and moving very locally. 

Especially we shall in our scheme of a 
fundamental 
string field theory, or shall we say Fock space, not have any explicit 
information on which little piece of string hangs together with which.
Such information shall rather be extracted - if it is at all possible -
by looking at whether there exist in our string field description 
a series of pieces that can make a considered pair of pieces hang 
together by making up some intermediate connection. It is characteristic 
for our scheme, that we make use of the well-known solution of 
the single string equations of motion by writing the 26-position field
variable $X^{\mu}(\sigma,\tau)$ as a sum of a right 
mover $X^{\mu}_R(\tau-\sigma)$ and a left mover $X^{\mu}_L(\tau+\sigma)$
part $\mu=0, 1, \cdots, 25$ (see equation(3.4) in section 3) and concentrate even on considering 
the derivatives with respect to $\tau$
of these two parts:
\begin{eqnarray*}
& \hbox{Our main variables}& \\*
& \dot{X^{\mu}}(\tau-\sigma) & \\*
& \dot{X^{\mu}}(\tau-\sigma) &
\end{eqnarray*} 

From these variables in a point on the string time track in 26 dimensional Minkowski space-time
we can at least obtain the direction of motion and the direction 
of pointing of an infinitesimal string bit. So the information of these
derivatives at least gives you the local motion and orientation of 
infinitesimal string bits from which one could then imagine to build up 
extended regions of a string time track. Only the information on 
the absolute position - which is though anyway dependent on the coordinate 
origo choice -, and the information on which among possibly many strings 
has the appropriate string time track (if we only give these derivatives but 
do not attach the ``name'' of the string) is not extractable from 
these derivatives. So apart from the distinction to say which strings 
have the various pairs of a right and a left mover derivative value
$(\dot{X}^{\mu}_R, \dot{X}^{\mu}_L)$ and apart from the absolute 
position - i.e. an additive constant in the position - the whole 
string development could be represented by such pairs 
$(\dot{X}^{\mu}_R, \dot{X}^{\mu}_L)$ if we got just the set of all 
such pairs.

Now, however, it must be admitted that we shall throw away even more 
information before fully obtaining OUR string field theory formalism:
We shall not even keep into our formalism the information as to which 
pairs $(\dot{X}^{\mu}_R, \dot{X}^{\mu}_L)$ occur, but ONLY keep the separate 
sets of $\dot{X}^{\mu}_R$ and  $\dot{X}^{\mu}_L$ which occur in these pairs.

This is a great simplification in as far as the reader should have in 
mind that respectively the right mover derivative  $\dot{X}^{\mu}_R$
and the left mover derivative $\dot{X}^{\mu}_L$ only depends on 
$\tau - \sigma$ and $\tau +\sigma$. Thus we achieve by this keeping 
only the components in the pairs $(\dot{X}^{\mu}_R, \dot{X}^{\mu}_L)$
that we reduce the kept information from the 
``two-dimensional'' manifold of pairs  $(\dot{X}^{\mu}_R, \dot{X}^{\mu}_L)$
to the only ``one-dimensional'' manifolds of single right 
or left mover derivatives $\dot{X}^{\mu}_R$ or $\dot{X}^{\mu}_L$.
However, this throwing out of information is not so serious again 
since it is largely given by continuity if you have only a few strings.
It is, however, not completely given by the continuity, but only mainly:
If we just know one pair $(\dot{X}^{\mu}_R, \dot{X}^{\mu}_L)$ 
to be present in the time track of some string and 
in addition which single right and left mover derivatives have 
values in the neighborhood of the respective values from the 
pair $(\dot{X}^{\mu}_R, \dot{X}^{\mu}_L)$, then from continuity you can 
find a neighborhood of pairs because you just combine all the neighboring 
single right and left mover derivatives. 

In this way we want to argue that just the information of which 
$\dot{X}^{\mu}_R$ and $\dot{X}^{\mu}_L$ there occurs on a set of strings 
gives by far the most information about these strings provided you 
make use of the continuity. Here we are thinking classically since 
quantum mechanically one would get worried about whether you 
can arrange a continuous string at all; there are namely in all field theories,
also the fields on the single string say, quantum fluctuations that 
make the field function 
strictly speaking non-continuous. But let us at least tentatively 
hope that there is some way of quantum mechanically also have some 
continuity left so that it can be used to recover some of the information 
we throw away.

\underline{The main idea of our model in the first approximation is now this:}

We represent a state of a string field theory system meaning a system 
of many strings, not by strings all, but by continuous curves formed 
by the set of all the values of  $\dot{X}^{\mu}_R$ and $\dot{X}^{\mu}_L$
found on any one of the strings (at any time). 

In the case of a string theory with only closed strings this would give
us two sets, namely one for the right mover derivatives and another one
for the left mover derivatives. So for an only closed string theory 
our model should in this first approximation be represented by two sets 
in two isomorphic but a priori different 26-Minkowski spaces. The set 
corresponding to the right mover derivatives is e.g.
 $I_R$ given by (4.8) below.

In the case of string theories with both open and closed strings it is 
more natural to only keep one set by uniting the two sets 
corresponding to right and left mover derivatives respectively. Since 
at the end of an open string the right mover waves are reflected as 
left movers and oppositely it would anyway turn out with open strings present
that the two sets constructed for right and left would become highly 
overlapping. So it would be a large amount of keeping the information 
double if we kept the two separate sets in the-with-open-string case.
Therefore we shall in \underline{the theory with open string case only} 
work with one set of derivatives of both right and left movers 
united.

It should be stressed again, that since we have thrown away information 
usually kept in string field theories we argue that our string field 
theory cannot be ``not new'' - unless compared to a candidate for a 
string field theory also throwing away information - and therefore 
it can only be possibly criticized that it throws away so much that it is no 
longer a string theory or that it is contradictory or not meaningful at all.

\subsection{The Second step in Constructing Our Model}

The obvious idea of the first approximation would now be to 
discretize the $\tau-\sigma = \tau_R$ and the 
$\tau+\sigma = \tau_L$ variables - much similar to the 
Thorn's discretizing the $\sigma$-variable \cite{new12}, but it is slightly different 
to discretize the right and left mover variables.
Then one would imagine to construct a Fock space in say with open string theory case corresponding to that there is a particle 
with a 26-momentum - essentially $\dot{X}^{\mu}_{\hbox{R or L}}$ -
for each time there is some string on which for some combination of 
$\sigma$ and $\tau$ the appropriate combination 
of the 26 components $\dot{X}^{\mu}_{\hbox{R or L}}$ occurs. If this were 
successful we would have transformed the information of the system of 
the several strings into a Fock space state, which a priori rather 
describes a series of particles, which in turn represents the derivatives of the 
right and left mover on the system of strings.

Now, however, it turns out that there is difficulty with this simple 
way: When one wants to construct a Fock space it is needed that 
the parameters to specify the single particle states into which 
we in the Fock space put some particles should commute among themselves.
For instance the usual way of making a Fock space is to use momentum 
eigenstates, and momenta commutes with themselves and each other 
so that it is a good way. One could also make a Fock space taking the outset 
from position eigenstate and that is how one basically obtain the 
$x$-dependent fields ${\bf \psi}(x)$, but one would get in trouble
having to set up a Fock space if one wanted to 
mix some components not commuting. 

Here our problem is that the derivatives of the right movers or just the right 
movers themselves do not commute in the single string description with 
themselves. Thus the $\dot{X}^{\mu}_{\hbox{R or L}}$ at one value of the 
argument $\tau_{R}  = {\tau} - {\sigma}$ and an infinitesimally neighboring 
value have non-trivial commutator or Poisont bracket and thus 
we cannot use the same parameters to make the Fock space for 
one value of $\dot{X}^{\mu}_{\hbox{R or L}}$ and for an infinitesimally 
close different value. Thus the simple project just mentioned does not quite 
work. 

To solve this problem we propose the idea of letting only the even 
points in the discretization of the variable $\tau_{R}  = \tau - \sigma$
be counted and put into the Fock space description. As we shall see below 
that would solve the problem provided we can interpret the discretized 
approximation to the well-known deltaprime commutator between the 
derivative of right or left  mover fields on the single string  
$\dot{X}^{\mu}_{\hbox{R or L}}$ (see formula (5.9) below) as meaning 
that a  $\dot{X}^{\mu}_{\hbox{R or L}}$-operator (in the single string theory)
which is discretized with an even number in the series of discretization 
points commutes with the other operators of this type except for the two
of course odd numbered nearest neighbors. You see we can claim that an
even discretization numbered  $\dot{X}^{\mu}_{\hbox{R or L}}$ commute with 
all other even ones. Thus there is no problem in building a 
Fock space up only based on the even ones alone.

But now luckily the odd ones are essentially just conjugate variables
to the even ones. It is a bit more complicated than that in as far 
as truly the odd  $\dot{X}^{\mu}_{\hbox{R or L}}$'s can be expressed as 
differences times an overall constant from the conjugate of the even ones.
This is seen in formula (1.2) how this relation is imagined 
provided one has in mind that $J^{\mu}$ is a discretized  
$\dot{X}^{\mu}_{\hbox{R or L}}$. But at least this thinking gives the 
hope that we might construct the odd discretization  
$\dot{X}^{\mu}_{\hbox{R or L}}$'s out of the conjugate variables of the even 
ones; and we claim
we shall see, that we did that successfully (at least locally on the string).
Like one in usual Fock space description can obtain positions of particles 
well enough even if one uses the momentum states to formulate the Fock space,
we can thus also expect - and we claim it works, - to get the odd 
$\dot{X}^{\mu}_{\hbox{R or L}}$'s represented even though we construct 
the Fock space only on the basis of the even discretized $\dot{X}^{\mu}_{\hbox{R or L}}$.

So you see we formulate our string field theory now as a Fock space 
ONLY for the even  $\dot{X}^{\mu}_{\hbox{R or L}}$'s -leaving the 
odd ones to come out as the conjugate analogous to having the position
come out as the conjugate if you start a Fock space in usual QFT 
on the basis of momentum eigenstates.

This was ``second approximation of the description starting from 
the strings'': We have reached to seeking to represent the state 
of a system of several strings by a discretized set of particles 
- we call them ``even objects'' - described by a Fock space. These 
particles (or ``even objects'') are then to be combined with their 
essentially conjugate ``odd objects''; and then all these ``objects''
represent the derivatives of the right or left mover fields on the 
strings   $\dot{X}^{\mu}_{\hbox{R or L}}$ in the ensemble of strings present.
It is already a slightly long reconstruction to get back to the string that our whole scheme can only be hoped 
to work if one can live without this little bit of lost information, first 
of all about which strings hang together how.

\subsection{String gauge choice and discretization}

But now we must also remember that we in string theory has the 
possibility of reparametrization and the problem of gauge choice.

We have in this article chosen what one calls light-cone gauge
which means that in a coordinate system with the metric 
(1.3) below one fixes the $\sigma$ coordinate such that 
the amount of $p^+$ momentum per unit $\sigma$ is constant.
We can even arrange that for the right and left mover parts separately 
because we are left over with a freedom to reparametrize right and left 
variables separately.
In addition we take as one almost always do in string theory 
from the beginning a choice of coordinates so that we have 
the simple Dalembertian in the two dimensional space time inside 
the string. Because of the original Nambu action one then get the 
constraints which in our concentration on the derivatives 
of the right and left movers take the especially elegant form
(3.5) and (3.6) below. 

Very naturally of course we imagine to take the discretization so
as to let the steps be equally long in the appropriate variable 
$\tau_R$ or $\tau_L$. 

When we now have to think of discretized $\tau_R$ or $\tau_L$ 
it becomes natural instead of working with the derivatives proper 
  $\dot{X}^{\mu}_{\hbox{R or L}}$ to use this derivative integrated up 
over the small interval associated with the discretization 
- the interval from half way to the to the left lying next 
discretization point to half way to the one to the right - 
and thus instead of directly working with $\dot{X}^{\mu}_{\hbox{R or L}}$
to use the integrals (5.1 -4)below, becoming then 
really differences between neighboring $X^{\mu}_R$ or $X^{\mu}_L$ 
variables (since integral of differentiation gives the difference).

The difference which is then really going to be our central variables
and which are to be used as the ``momentum variable'' in constructing 
the Fock space is called $J^{\mu}_R$ or $J^{\mu}_L$ for respectively 
right and left mover - while in the case with open string too we
mix them together to say just $J^{\mu}$ - is then given as say
\begin{eqnarray*}
J^{\mu}(I)  = X^{\mu}_{\hbox{R or L}}(I+1/2) -X^{\mu}_{\hbox{R or L}}(I-1/2)
\end{eqnarray*} 
which will appear in section.5 as equations (\ref{5.3}) and (\ref{5.4}).
Here we have written a discretization number counter $I$ and 
we even use the non-integer value for it to specify the 
middle points between the discrete points on the variable 
$\tau_{\hbox {R or L }}$ axis in question. 

When we in foregoing subsection talked about that we used the 
 $\dot{X}^{\mu}_{\hbox{R or L}}$ to give the ``momenta'' for which there 
should be a particle in the Fock-space theory describing at the end 
our string theory, that should be replaced by using these 
$J^{\mu}$ variables instead of the  $\dot{X}^{\mu}_{\hbox{R or L}}$'s.

So finally we arrive at proposing the $J^{\mu}$ to be used to construct 
the Fock space. But have still in mind that we only use the 
``even ones'' of these $J^{\mu}$'s of course since we already only 
used as described above the even ones of the  $\dot{X}^{\mu}_{\hbox{R or L}}$'s.

Now we however have the constraints (3.5) and (3.6) which trivially 
can approximately be rewritten to be about the $J^{\mu}$'s. In fact we get 
(5.8).

In the spirit of the light cone gauge we shall use that the plus component $J^+$
of the $J^{\mu}$ gets fixed, and the minus component $J^{-}$ can be solved for 
using the constraint (5.8). Thus the independent degrees of freedom are only 
the ``transverse'' components corresponding to the remaining 26 -2 = 24 
dimensions $J^i$ (we use the notation $i$ for these 
24 transverse components).

So it ended up that a state of the system of strings is 
in a not hundred percent complete way - but almost fully -
represented by a Fock state in which we have ``particles'' 
(or we call them ``objects'') each of which represents an even 
discretized $J^{\mu}$ but it is only needed to give the set
of transverse components $J^i$ ( i = 1,2 ,3,...,23,24.).

(Since we are working with bosonic strings we denote the dimension,
even if never get far enough to see much of the significance for that,
as 26 =25 +1.).

\subsection{How scattering has been removed}

With our here described removal of information in going from the string picture to the picture of our model we shall see that we removed so much information that we have in our model nothing happening fundamentally under the scattering. In fact we review in section 4 below that the information which we kept in our model, namely the set of values for $\dot{X}^{\mu}_{R}$ and $\dot{X}_{L}$, called there $I_{R}$ and $I_{L}$, is \underline{not} changed during the scattering. This means that our string field theory, considered a theory of the ${J}^{\mu}$ essentially $\dot{X}^{\mu}_{R \ or \ L}$ has no scattering. So we should rather say that our string field theory is built under the following point of view:

Our point of view is that string field theory (meaning full string theory with the possibility of many strings around in the description) taken in our formalism is essentially a theory without any genuine interactions. Indeed we want in the following to argue for that our string field theory is solvable and that the scatterings \underline{at the end} described by the Veneziano model(s) are - at least in some sense - \underline{faked}. That is to say in the formalism, which we use below (and in our foregoing articles \cite{1}, \cite{2}) to describe the string field theory, nothing really happens - even when looked upon as a theory of strings -, although these strings scatter on each other. This statement may sound strange or contradictory. But we shall explain in the sections below how it is possible.

Our string field theory is actually best put forward as an ``ontological'' setup that at first has nothing to do with strings, but is rather described by what we call ``objects''. These ``objects'' are more like particles than strings in as far as each ``object'' has - when we aim at the bosonic string theory with ideally $25 + 1$ dimensions - $24 \ J^{i}$ variables and in addition $24 \ \Pi ^{i}$ conjugate momenta to the $J^{i}$'s for that object. It is o.k. in our formalism at first to think of one of our ``objects'' as a system/a particle with $24$ position components $J^{i}$ and $24$ (conjugate) momenta $\Pi^{i}$ to them. The ``ontological'' model to be at the end by a relatively complicated rewriting described as a string field theory consists of an arbitrary (it is part of the Fock space information how many) number of ``objects'' and is simply a second quantized system of these ``objects'' (which as mentioned before with their only $24$ degrees of freedom are more remi
 niscent of particles than of strings at first). That is to say we shall have for every point in the space with $24$ coordinates identified with $J^{i}(i = 1, 2, \cdots, 24)$ a creation $a^{+}J^{i}$ and annihilation $a(J^{i})$ operator creating or annihilating an ``object'' with just that $J^{i}$ vector.

In order to proceed from these ``objects'' to their interpretation to describe strings you first imagine them organized into closed series of objects by picking out of the set of objects cyclically ordered series of objects. Let us enumerate the objects selected to be such a cyclically ordered series by an integer $I$, of which a technical detail will be presented in a moment, is supposed to run through the \underline{even} integers only, 

\begin{eqnarray}
\left.
\begin{array}{c}
 J^{i}(I) \\
 \Pi^{i}(I)  
\end{array}
\right\}
 \mbox{with $I=0,2,4,\dots, M-4, M-2$}
\end{eqnarray}
where $M$ is some even number equal to twice the number of ``even objects'' in the cyclically ordered series considered.

We only consider the ``(even) objects'' here as the fundamental ones, but we may sometimes also denote some ``(odd) objects'' which are genuinely constructed from the canonically conjugate variables to the ``(even) objects'' as ``objects'' although they are not a priori in the ``fundamental'' string field theory model.

\begin{figure}[!htb]
\begin{center}
 \includegraphics[width=0.5\textwidth]{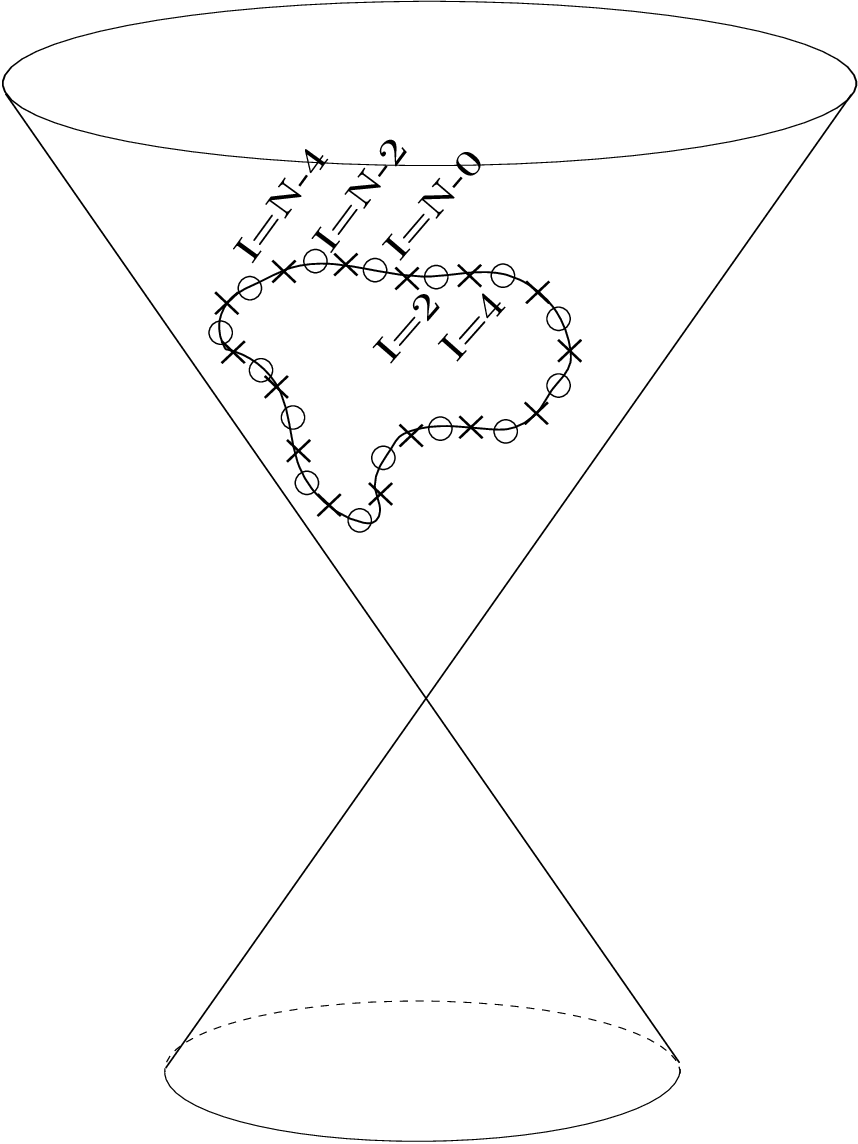}%
\end{center}
 \caption{Illustration of a cyclically ordered chain of ``objects'' with the even ``objects'' 
          denoted as small circles $\circ$ and the odd ``objects'' 
          as small crosses $\times$. The cycle is symbolized by a curve ----.\label{Figure2}}
 \end{figure}

The point of this enumerating the closed chain of ``objects'' by even numbers is that the next step in going along to connect our formalism with strings is to ``invent'' or rather construct just formally a series of further ``objects'' that go in on the odd places, being assigned an odd $I$. In fact we construct such an ``added'' or ``invented'' the closed chain of the object with an odd number $I$ marked ``object'' by means of the conjugate variables $\Pi ^{i}$ of in the closed chain neighboring even ``objects'' by the definitional formula for $I$ odd case:

\begin{eqnarray}
J^{i}(I)=-\pi \alpha ' 
 \left( \Pi^{i}(I+1)- \Pi^{i}(I-1) \right) \ 
 \ \ \mbox{for \ odd \ } I.
  \label{odd}
\end{eqnarray}

 (Notice that the string tension $\frac{1}{2 \pi \alpha'}$ is here ``sneaked'' in in a way only to be understood later.)
 
The next step in constructing the string field theory is to formally extend the formalism a ``$25+1$ dimensional space-time'' for the objects by again inventing such that each object having at first only $24$ components $J^{i}(I)$ should have two more components so as to get to $25+1$ ($25$ space and one time). Using the IMF (= infinite momentum frame) metric tensor
 
\begin{eqnarray}
g_{\mu\nu}= 
\begin{array}{c}
+ - \ \ \overbrace{ \hspace{6em}}^{24}\\
\left(
 \begin{array}{cccccc}
 0 & 1 & 0 &\cdots & \cdots& 0\\
 1 & 0 & 0 & \cdots & \cdots & 0\\
 0 & 0 & -1 & \cdots & \cdots & 0\\
 \vdots & \vdots & & \ddots & & \vdots\\
 0 & 0& \cdots &\cdots& -1 & 0\\
 0 & 0 & \cdots & \cdots & 0 & -1
 \end{array}
\right)
\end{array}
\begin{array}{c}
 \\ + \\ - \\ 
   \left. \begin{array}{l}
    \\ \\ \\ \\    
   \end{array}
    \right\} 24
\end{array}
\label{metric}
\end{eqnarray}
and fixing the +components of all the ``objects'' even $I$ as well as the invented odd $I$ ones to be a fixed number denoted

\begin{eqnarray}
J^{+}(I)= \frac{a \alpha'}{2}\left( = \frac{2 \pi \alpha' a}{4 \pi} \right)
\label{gf}
\end{eqnarray}
and adjusting the $J^{-}(I)$ components to make (25+1) - vector $J^{\mu}(I)$ of an ``object'' be light-like we obtain the condition
\begin{eqnarray}
2J^{+}(I)J^{-}(I)-\sum_{i=1}^{24} \left( J^{i}(I) \right) ^{2}= 0
\label{ll}
\end{eqnarray}
In the expression (\ref{gf}), \underline{$\alpha'$ denotes the Regge slope} and \underline{"$a$" is lattice distance} when we introduce for $\tau_ R(I)$ and $\tau_ L(I)$, discretization in section 5.
That is to say we have just introduced by definition
\begin{eqnarray}
J^{-}(I)\stackrel{\wedge}{=} \frac{\sum_{i=1}^{24}\left( J^{i}(I)\right) ^{2}}{2 J^{+}(I)}
=\frac{\sum_{i=1}^{24}\left( J^{i}(I)\right)^{2}}{a \alpha '}
\label{jm}
\end{eqnarray}
the $J^{-}(I)$ component for any  ``object'' like the $J^{+}(I)=\frac{a \alpha '}{2}$.

Together with the $24 J^{i}(I)$ components of an ``object'' number $I$ the two extra components 
\begin{eqnarray}
J^{-}(I)=
\left\{
\begin{array}{c}
\frac{\sum_{i=1}^{24} \left( \Pi^{i}(I+1)- \Pi^{i}(I-1) \right) ^{2}\pi^{2}\alpha'}{a}
\ \ \ \  \mbox{for \ I \ odd \ }\\
\frac{\sum_{i=1}^{24}\left(J^{i}(I) \right) ^{2}}{a \alpha}
\ \ \ \ \ \ \ \ \ \ \ \ \ \ \ \ \ \ \ \ \mbox{for \ I \ even}
\end{array}
\right. 
\end{eqnarray}
and 
\begin{eqnarray}
J^{+}(I)=\frac{a \alpha '}{2}
\end{eqnarray}
make up a light-like $25+1$ dimensional Minkowski space-time $26$-vector.

Apart from an integration constant - which is ambiguous - we may sum up along one of the cyclically ordered chains stating from some starting point chosen to obtain a sum
\begin{eqnarray}
X^{\mu}_{R}\left( I+\frac{1}{2} \right) = \sum_{\mathrm{start ~point~ of~ I}}^{I}J^{\mu}(I)
\label{sum}
\end{eqnarray}
which we want to identify with the $X_{R}(\tau_{R})$ of a string in the state of several strings. More precisely and generally we say at first that there is a ``potential'' string passing through a given space-time point in a $25+1$ dimensional Minkowski space-time $X^{\mu}$, provided there two such sums as (\ref{sum}) and that 
\begin{eqnarray}
X^{\mu}=X^{\mu}_{R}\left(K+\frac{1}{2}\right)+X^{\mu}_{R}\left(I+\frac{1}{2}\right).
\label{s}
\end{eqnarray}
You should think of this as meaning that the right and left moving parts of the single string solution to the equations of motion 
\begin{eqnarray}
X^{\mu}(\sigma, \tau)=X^{\mu}_{R}(\tau - \sigma) + X^{\mu}_{L}(\tau + \sigma)
\end{eqnarray}
are identified with the sums of the type (\ref{sum}) of the $J^{\mu}(I)$ from some starting point $I_{\mbox{start}}$. Then when we as here consider - for simplicity - a theory with open strings we actually have reflection of the waves on the string at the end points of string and there actually get the boundary condition, 
\begin{eqnarray}
\dot{X}^{\mu}_{R}(\sigma =0, \tau)=\dot{X}^{\mu}_{L}(\sigma=0, \tau)
\end{eqnarray}
so that for open strings we do not have to distinguish right and left movers (provided we have the correct periodicity in $\sigma$ imposed).

Now, however, it should be said that if we mix together sums of over $J^{\mu}(I)$'s of the type (\ref{sum}) from different cyclically ordered chains and construct the sum of two terms (\ref{s}) then we get the ``potential'' space-time points through which a string \underline{may} pass. However, it is suggested that it is only a tiny subset of these ``potential'' space-time points which should truly be thought to have a string passing through them. However, we shall normally imagine that constructing a term of two sums of the type (\ref{sum}) from the same cyclically ordered chain would give a space-time point $X^{\mu}$ through which indeed a string passes, namely the open string associated with the cyclically ordered chain of  ``objects'' in question.

As one can see the relation between our basic description in terms of evenly numbered ``objects'' in 24 dimensions to the string description in the $25+1$ dimensions involves quite a few steps: a) introduction of odd ``objects'' constructed from the conjugate momenta $\Pi^{i}(I)$, b) introduction of two extra dimensions, c) interpretation of the string points as a sum of two respectively right and left mover terms, d) that only part of the ``potential'' string space-time points are truly representing passage of a string.

So it is a bit complicated relation.

This in reality means that the string field theory we construct is rather much a mathematical construction from our ``objects'' in $24$ dimensions which are at first looking like having nothing to do with string theory.

We should therefore rather think of our string field theory as a mathematical construction which solves string theory with several strings in terms of a mathematical language that just happens to be our ``objects''.

We want in the present article to argue for that the model or string field theory of ours have several of the properties of a wanted string theory (of several strings).

The main progress of the present article compared to our earlier articles \cite{1}, \cite{2} is that we here deduce the spectrum of the strings in our model. That is to say we deduce the fact that the strings extracted from our formalism have the completely usual mass square spectrum as the single string - apart from some small caveats: a species doubler and some other discretization effects appear, both shown to disappear in the continuum limit, meaning $a \to 0$. Having in mind usual string field theories such as \cite{3}-\cite{11}, wherein the single string propagation is built in from the start into the setup, it would seem that in the limit of weak coupling it has been basically put in, that the single string spectrum must be the wellknown single string spectrum. Thus it would not seem an impressive progress to show that the spectrum is indeed the usual single string spectrum. However, in our model we have \emph{no truly built in single string propagation} and thus it be
 comes less trivial, that our model has the expected spectrum. We have roughly speaking \emph{dissolved} the single string degrees of freedom into separate left and right movers (in the closed string theories totally, in the open string theory only locally), so that there are \emph{no strings} in our ``ontological'' formalism directly; the single string theories rather only come in via at least not totally trivial mathematical constructions gotten from the ``ontological'' setup by the just described couple of introductions of new variables. Obtaining the mass spectrum of the single string thus checks, that our mathematical definitions of further variables applied to the ``ontological'' setup indeed leads well enough to the string theory, that at least the spectrum of the single string gets inherited into our model. So it is at least no more trivial than our whole construction of the new degrees of freedom and our ``ontological model'', since it is in fact one - and essentiall
 y the first detailed one - of the tests to be performed on our formalism to ensure, that indeed it is the string theory. Otherwise it could be that it is only a suggestive dream that our model constitutes string theory. We still think that other tests are needed, but consider it psychologically a very promising result, that the spectrum seems as we shall see below, to have come out right.

The very fact that we allow for more ``objects'' than the ones in one cyclically ordered chain means that there can be arbitrary many such chains and thus strings. So in this sense we have an arbitrary number of chains and thus strings so that we have in principle the string field theory.

Since in our ``objects'' formalism there is no time development it may seem very surprising if we can indeed get such a formalism to describe scattering of strings. In the following section \ref{section 2} we shall argue for that it is indeed not impossible that a theory without any time development (zero Hamiltonian so to speak) can at all describe scattering. In section \ref{section 3} we shall remind about the formalism of usual single string theory and in section \ref{section 4} review shortly the conservation law of the images of the left and right mover $\dot{X}_{L}$ and $\dot{X}_{R}$ which is so crucial for our string field theory and which formally show that (classical) string theory is ``solvable''. Then in section \ref{section 5} we return to the described system of ``objects''in the introduction. In section \ref{section 6} we then put forward one of the (great) achievements of our formalism: the spectrum of string states.

In section \ref{section 7} we start on the ideas for deriving the Veneziano model for scattering of strings in our string field theory, but we shall be satisfied with a setup and postpone the full derivation of the Veneziano model \cite{21} to our later publication \cite{22}.

In section \ref{section 8} we resume and conclude.

\section{Scattering without anything happening \label{section 2}}  

Let us have in mind that the Hilbert space of possible state of the world of a string theory in our SFT model is in fact the Fock space for what we call ``objects''. These ``objects'' are each essentially a particle or a system with $24$ degrees of freedom, meaning $24 \ J^{i}(I)$-variables and $24 \ \Pi^{i}(I)$- variables canonically conjugate to the $J^{i}$'s.

Even when these ``objects'' in a slightly complicated way describe scattering of strings they themselves do \underline{not} develop. In our ``object'' formulation everything is totally static, or rather there is no time, it is timeless. 
 
This scattering without anything changing sounds a priori very strange. Therefore we would like here to give at least an idea of how that strange phenomenon can come about:

Suppose that we had a couple of series of constituent particles making up some composite particles (essentially bound states, but they might not even be bound; rather just formally considered composed). Now if one decides to divide the ``constituent'' particles into groups forming composites in a different way from at first, then the momenta of the composite clumps after the considering, the new ones, will typically be quite different from those of the initial composite clumps.

\begin{figure}[!htb]
\begin{center}
\includegraphics[width=0.35\textwidth]{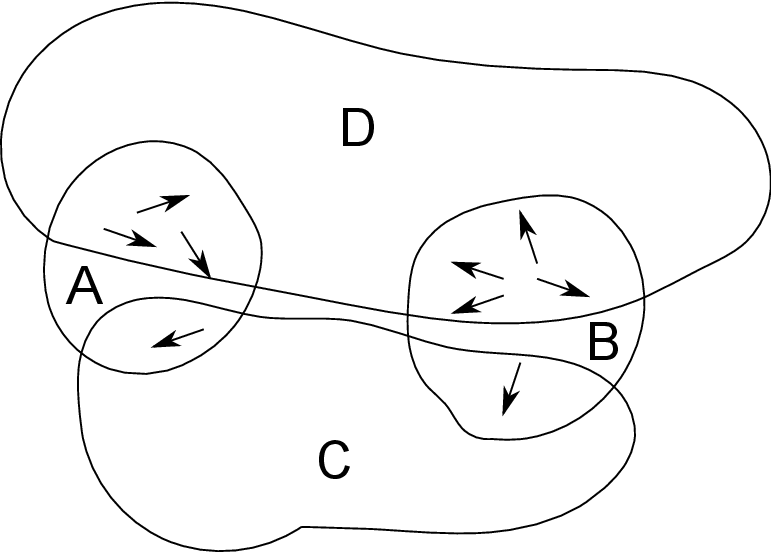}%
\end{center}
\caption{We illustrate how one may look at a set of (independent) ``constituents'' forming first two clumps $A$ and $B$, while later we divide them into two clumps $C$ and $D$ in a different way. \label{Figure3}}
\end{figure}

\begin{figure}[!htb]
\begin{center}
 \includegraphics[width=0.35\textwidth]{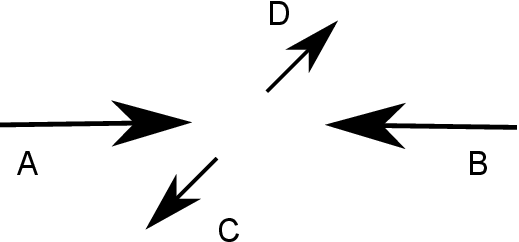}%
\end{center}
  \caption{Counting the momenta of the clumps $A, B, C, D$ of the foregoing Figure  \ref{Figure3} as the sum of the momenta of the ``constituents'' 
           we get the picture of a \underline{scattering} $A+B \to C+D$.\label{Figure4}}
\end{figure}

This is a quite trivial remark: If we reclassify some constituents into a new set of classes of constituents (i.e. new composites) of course it will look like scattering of the composites.

In this way we hope that the reader can see that there is also a chance that making up a model on our ``objects'': these ``objects'' can function much like the just mentioned ``constituents'' and thus we could also in our model have some pretty formal scatterings. In the way we here think of the scatterings these scatterings are something we only think upon. The constituents and analogously our ``objects'' do not change their momenta say at all.  We have in this sense presented the idea of having completely formal scattering without anything going on at all for our ``objects''. It is our hope in long run to argue that in spite of this scattering being very formal we shall at the end get for it scattering amplitudes becoming the Veneziano model amplitudes.

\section{Reminding String Theory} \label{section 3} 

In the introduction we told that our string field theory is described by a Fock space of what we called ``objects'' and that the ``objects'' have $24$ degrees of freedom each. In order for the reader to connect this ontological model - one could say - of ``objects'' to strings it may be useful to have in mind some well-known facts about string theory:

After having chosen what is called conformal gauge - in say the Nambu action - string model we obtain for the $X^{\mu}(\sigma. \tau)$ variables on the string space-time track the equations of motion
\begin{eqnarray}
(\partial_{\tau}-\partial_{\sigma})(\partial_{\tau}+\partial_{\sigma})X^{\mu}=0
\end{eqnarray}
where $\tau$ and $\sigma$ are the ``time'' and ``space'' coordinates on the string, and the constraints are 
\begin{eqnarray}
\dot{X}^{\mu 2}-(X'^{\mu})^{2}&=&0 \\
X^{\mu '}\cdot X'_{\mu}&=&0.
\end{eqnarray}

These equations \cite{17}, \cite{18} are solved in terms of right $X^{\mu}_{R}(\tau-\sigma)$ and left $X^{\mu}_{L}(\tau+\sigma)$ mover fields by 
\begin{eqnarray}
X^{\mu}(\sigma, \tau)=X^{\mu}_{R}(\tau-\sigma)+X^{\mu}(\tau+\sigma)
\end{eqnarray}
and the constraints are given in terms of right and left movers separately:
\begin{eqnarray}
\left( \frac{dX^{\mu}_{R}(\tau_{R})}{d \tau_{R}} \right)^2&=& 0 \\
\left( \frac{dX^{\mu}_{L}(\tau_{L})}{d \tau_{L}} \right)^2&=& 0
\end{eqnarray}
where we put $\tau_{R}=\tau-\sigma$ and $\tau_{L}=\sigma+\tau$.

It is the choice in our model essentially to concentrate on the derivatives with respect to the relevant variables $\tau_{R}=\tau-\sigma$ or $\tau_{L}=\tau+\sigma$ of right and left mover fields
\begin{eqnarray}
\dot{X}(\sigma, \tau)=\dot{X}^{\mu}_{R}(\tau-\sigma)+\dot{X}^{\mu}_{L}(\tau+\sigma)
\end{eqnarray}
and the constraints
\begin{eqnarray}
\left( \dot{X}^{\mu}_{R}(\tau_{R}) \right) ^{2}=0=\left( \dot{X}^{\mu}_{L}(\tau_{L}) \right) ^{2}
\label{constraint}
\end{eqnarray}
meaning that these derivatives $\dot{X}_{R}(\tau_{R})$ and $\dot{X}_{L}(\tau_{L})$ represent mappings of single variables $\tau_{R}=\tau - \sigma$ and $\tau_{L}=\sigma + \tau$ into the light-cone $25+1$ dimensional space. The constraints namely mean that these derivatives lie on the light-cones.

\section{The constant Images $I_{R}$ and $I_{L}$ or for open case $I_{R} \cup I_{L}$} \label{section 4} 

We have already a couple of times published a theorem \cite{1}, \cite{2} which holds for classical string scattering and which is in some way the basis for our presently discussed string field theory.

Our theorem concerns strings with the same properties as in usual string theory meaning e.g. that they are described in \cite{17}, \cite{18} by the Nambu-action \cite{13}, \cite{14}, only we describe them classically, i.e. without quantum mechanics. We take their scattering to be like ``u-channel scattering'' in the sense that we imagine two different strings to touch each other in a point at some moment of time, and then they so to speak exchange tails, as is depicted in Figure  \ref{Figure1}.

There is in the very touching moment 4 pieces of string going in or out from the same point. At the start two of these pieces of strings make up one string and the two other make up the other string. After the ``exchange of tails'' the 4 pieces get put together into strings in a new way.

Now our theorem concerns images of the functions $\dot{X}^{\mu}_{R}(\sigma, \tau)= \dot{X}^{\mu}_{R}(\tau- \sigma)$ and $\dot{X}^{\mu}_{L}(\sigma, \tau)= \nobreak \dot{X}^{\mu}_{L}(\tau- \sigma)$ meaning  the set of values taken on by these functions, which when we think of \underline{closed} strings, are 
\begin{eqnarray}
I_{R}=\bigcup _{ {s=1,\cdots,0}  \atop {\mathrm{ for}~ \mathrm{string}~ s }}I_{Rs}
\end{eqnarray}
where
\begin{eqnarray}
I_{Rs}= \left\{ \dot{X}^{\mu}_{R}(\tau-\sigma)|(\sigma \mbox{coordinate on string} \ s \right\}
\end{eqnarray}
and
\begin{eqnarray}
&&I_{L}=\bigcup_{{s} \atop s~\mathrm{are~string}}
I_{Ls} \\
&&I_{Ls}(\tau)=\{ \dot{X}_{L}(\tau+\sigma)| \ \ 0 \leq \sigma \leq \pi \ \mbox{coordinates for string} \ s \}
\end{eqnarray}
Here $\dot{X}^{\mu}_{R}$ and $\dot{X}^{\mu}_{L}$ are the derivatives of the right and left mover fields $X^{\mu}_{R}$ and $X^{\mu}_{L}$ in terms of which the equations of motion in conformal gauge of the string $X^{\mu}(\sigma, \tau)$ is solved
\begin{eqnarray}
X^{\mu}(\sigma,\tau)=X^{\mu}_{R}(\tau-\sigma)+X^{\mu}_{L}(\tau+\sigma).
\end{eqnarray}
As time, which we may essentially identify with $\tau$, goes on the images $I_{Ls}(\tau)$ do \underline{not} change, because a shift in $\tau$ can be compensated by a shift in $\sigma$ and the image $I_{Rs}$ is the set of all the values taken for different $\sigma$-values. It is at best so that in the intervals between scatterings - i.e. between the moments when some of the string touches each other - the images $I_{Rs}(\tau)$ and $I_{Ls}(\tau)$ do not change. Thus we do not need to write the dependence on $\tau$ since $I_{Rs}(\tau)$ and $I_{Ls}(\tau)$ are constant as function of $\tau$ between the scatterings. When a scattering takes place we obtain a new set of strings and a given string $s$ stops to exist. 

Now the crucial point of our theorem is that even when a classical scattering ($\sim$ exchange of tails) takes place the unifications
\begin{eqnarray}
I_{L}=\bigcup_{s \ {\mathrm{strings}~}}I_{Ls}
\end{eqnarray}
and
\begin{eqnarray}
I_{R}=\bigcup_{s \ \mathrm{strings}~}I_{Rs}
\end{eqnarray}
do \underline{not} change during the scatterings. Since the $I_{Ls}$ and $I_{Rs}$ do not change between scatterings it means that these unifications $I_{L}$ and $I_{R}$ are totally conserved in \underline{closed} string theories. That there cannot be any change in these unions $I_{R}$, $I_{L}$ in the very scattering moments - except for a null set possibly - is understandable by thinking of the system of strings in such a scattering moment as composed by pieced bounded by the touching point. 

\begin{figure}[!htb]
\begin{center}
  \includegraphics[width=0.5\textwidth]{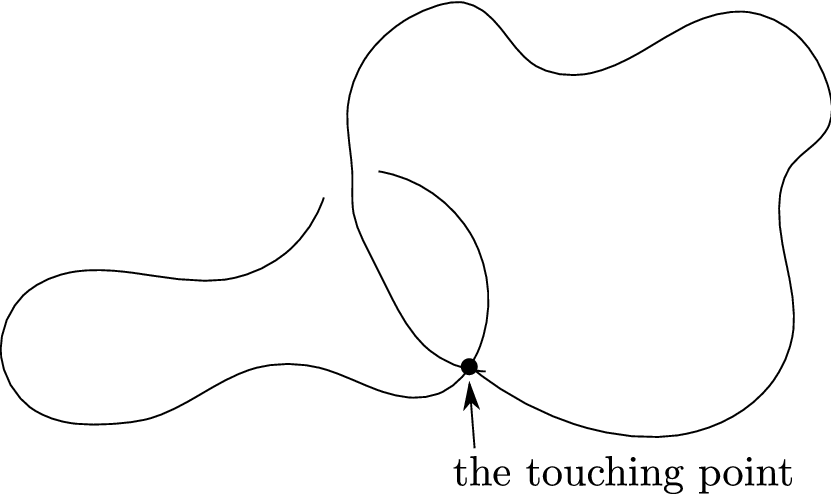}
\end{center}
   \caption{Two closed strings in the very moment of touching. In the scattering they become one long closed string 
             almost touching itself immediately after the scattering moment. \label{Figure5}}
\end{figure}

\begin{figure}[!htb]
\begin{center}
  \includegraphics[width=0.5\textwidth]{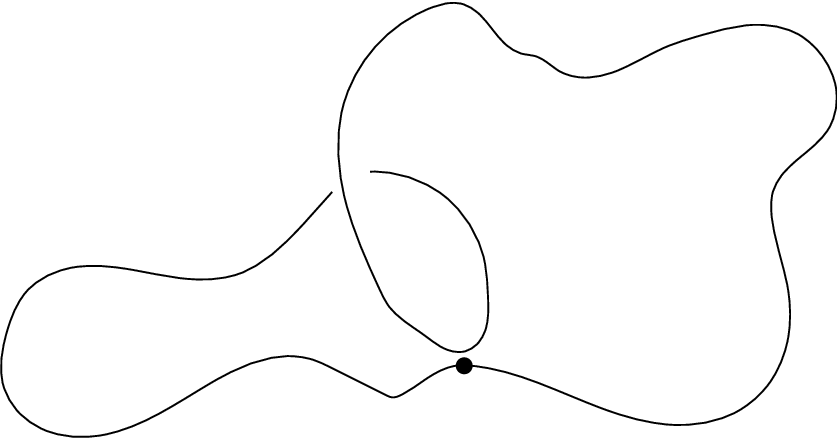}%
\end{center}
 \caption{Infinitesimally shortly after a scattering of two ``classical'' closed string the picture of the string is still close to 
          the picture of the two strings that touched, but now there is only \underline{one} (closed) string of an ``object'' with the neighbors to both sides.
           \label{Figure6}}
\end{figure}

From locally so to speak nothing can happen due to the scattering away from the very touching point at the very moment of scattering-touching. Especially the $\dot{X}^{\mu}_{R}$ and $\dot{X}^{\mu}_{L}$ away from the touching point(s) cannot change. So even less can the image of these $\dot{X}_{R}$ and $\dot{X}_{L}$ change and thus $I_{L}$ and $I_{R}$ must be unchanged during the infinitesimal time of touching and as we have seen also in between the scatterings.

This gives our theorem for closed string theories:

\underline{Theorem for closed strings}:

Under scattering (with piece exchange interactions) and time development of \underline{classical} dual strings - derived from say Nambu action - the union of the right mover derivatives 
\begin{eqnarray}
I_{R}=\bigcup_{s \ \mathrm{strings}~}I_{Rs}
\end{eqnarray}
where
\begin{eqnarray}
I_{Rs}=\left\{ \dot{X}_{R}(\tau-\sigma)| {\sigma \ \mbox{on string} \ s} \right\}
\end{eqnarray}
for all the strings $s$ remain a constant set on the light-cone in $25+1$ dimensional Minkowski space. (up to a null set).

Similarly the union $I_{L}=\cup I _{\stackrel{\mathrm{string} s}{Ls}}$ the left mover derivative images over all the strings $s$ remain forever a constant set modulo a null set on the left cone in an other $(25+1)$ dimensional Minkowski space-time.

While we in theories of only closed strings get a separate conserved set for right- and left- movers, we have for open strings boundary conditions that at the end points where right mover fields on the string get reflected as left mover fields and oppositely. Indeed we have e.g. at $\sigma=0$ the condition
\begin{eqnarray}
X'^{\mu}(\sigma=0,\tau)=0
\end{eqnarray}
which implies
\begin{eqnarray}
\dot{X}^{\mu}_{R}(\tau-\sigma)=\dot{X}^{\mu}_{L}(\tau+\sigma) \ (\mbox{at} \ \sigma=0)
\end{eqnarray}
meaning
\begin{eqnarray}
\dot{X}^{\mu}_{R}(\tau)=\dot{X}^{\mu}_{L}(\tau).
\end{eqnarray}

Thus for theories with open strings we get the right and left movers ``mixed up'' and it is not hard to see that it is in this case rather the union $I_{O}=I_{R} \cup I_{L}$ of both right- and left- mover derivatives that becomes conserved than the two sets separately: 

\underline{Theorem with open strings}

Under scattering and time development of a system, of strings containing also open strings and treated classically and with ``tail exchange'' interactions the combined image of both left- and right- mover derivatives $I_{O}=I_{R}\cup I_{L}$ for all the strings is conserved. 

These theorems in themselves mean that string theory even with scattering should be classically (and essentially) solvable, because these unions
\begin{eqnarray}
I_{R}, I_{L} \ \mbox{and} \  I_{O}=I_{R}\cup I_{L}
\end{eqnarray}
being conserved means that there are enough conservation laws that we must consider the string theory even including scattering a ``solvable theory''.

These theorems make up the basis for our string field theory in as far as our string field theory can be considered an attempt to formulate a description of a system of arbitrarily many strings ($\sim$ a string filed theory) by making use of precisely these according to our theorem(s) considered conserved sets $I_{O}, I_{R}, I_{L}$.

\section{The ``objects'' formalism} \label{section 5} 

In the light of the theorem in foregoing section that in the closed string case $I_{R}$ and $I_{L}$ (while for open string models we have only one conserved image $I_{R}(=I_{L})$) we see the idea of putting up a completely stationary theory by making a description in terms of these images the basis of the description.
Also in the open string case we have only one conserved image $I_{O}$ and in the same manner with the closed string case we may perform description in terms of the image $I_{R}$ and $I_{L}$. 

We want to discretize the description of $\dot{X}^{\mu}_{R}$ and $\dot{X}^{\mu}_{L}$ by defining (see last year's (2011) Bled proceedings \cite{2})
\begin{eqnarray}
J^{\mu}_{R}(I)=\int^{\frac{1}{2}(\tau_{R}(I)+\tau_{R}(I+1))}_{\frac{1}{2}(\tau_{R}(I)+)\tau_{R}(I-1)} \dot{X}^{\mu}_{R}(\tau_{R})d\tau_{R}
\end{eqnarray}
and
\begin{eqnarray}
J^{\mu}_{L}(I)=\int^{\frac{1}{2}(\tau_{L}(I)+\tau_{L}(I+1))}_{\frac{1}{2}(\tau_{L}(I)+)\tau_{L}(I-1)} \dot{X}^{\mu}_{L}(\tau_{L})d\tau_{L}
\end{eqnarray}

Here we have put a series of discretized points $\tau_{R}(I)$ and $\tau_{L}(I)$ on the $\tau_{R^{-}}$ and $\tau_{L^{-}}$ axes respectively in a dense way. The integer $I$ is one enumerating the steps in these discretizations. The $25+1$ vectors $J^{\mu}_{R}(I)$ and $J^{\mu}_{L}(I)$ are integrals over the discretization steps of respectively $\dot{X}^{\mu}_{R}(\tau)_{R}$ and $\dot{X}^{\mu}_{L}(\tau)_{L}$ or equivalently they are differences
\begin{eqnarray}
J^{\mu}_{R}(I)\simeq X^{\mu}_{R}\left( \tau_{R}(I+ \frac{1}{2}) \right) - X^{\mu}_{R}\left( \tau_{R}(I- \frac{1}{2}) \right)
\label{5.3}
\end{eqnarray}
and
\begin{eqnarray}
J^{\mu}_{L}(I)\simeq X^{\mu}_{L}\left( \tau_{L}(I+ \frac{1}{2}) \right) - X^{\mu}_{L}\left( \tau_{L}(I- \frac{1}{2}) \right)
\label{5.4}
\end{eqnarray}

In the case of open strings when we have a boundary condition at $\sigma = 0$ saying 
\begin{eqnarray}
X^{\mu}_{L}(\tau)=X^{\mu}_{R}(\tau)
\end{eqnarray}
we shall not distinguish $R$ and $L$ but rather define just one
\begin{eqnarray}
J^{\mu}(I)=X^{\mu}_{R \ \mathrm{or} \ L}\left( \tau_{R \ \mathrm{or} \ L}\left(I+ \frac{1}{2}\right) \right) - X^{\mu}_{R \ \mathrm{or} \ L}\left( \tau_{R \ \mathrm{or} \ L}\left(I- \frac{1}{2}\right)\right)
\label{diffBl55}
\end{eqnarray}

It is the main point of our model or string field theory to ``liberate'' these integrals of the $\dot{X}^{\mu}_{R}(\tau_{R})$ and $\dot{X}^{\mu}_{L}(\tau_{L})$ (for open strings $\dot{X}^{\mu}_{L}(\tau) = \dot{X}^{\mu}_{R}(\tau)$) which according to our theorem are unchanged even under scattering. This \underline{main idea} is that apart from an unimportant null set all the information about the state of the string field theory universe is contained in the totally conserved $I_{R}$ and $I_{L}$ (or in the open string case the union $I_{R} \cup I_{L}$). Then discretized this information about the universe state is contained in the set of values for the $(25+1)$-vectors $J^{\mu}_{R}(I)$, $J^{\mu}_{L}(I)$ (or in the open string case just $J^{\mu}(I)$).

Since we noticed in the discussion of our theorem that these sets $I_{R}$, $I_{L}$ or in open case $I_{R} \cup I_{L}$ keep the same elements even during scattering when some of these elements go from one string to another one, we should not consider the $\dot{X}^{\mu}_{R}(\tau_{R})$ and $\dot{X}^{\mu}_{L}(\tau_{L})$ images or between the ``objects'' $J^{\mu}_{R}(I)$ and $J^{\mu}_{L}(I)$ as forever associated with given strings, but rather as \underline{liberated} to be independent ``objects''. In the case of having open strings e.g. we have only one type of ``objects'' and they are just denoted as $J^{\mu}(I)$ depending on some discrete enumeration.

Thus our main strategem is to build a theory in a completely static way from what we call ``objects'' $J^{\mu}(I)$. That is to say we imagine a (discrete) set of so called objects each of which are a system of - from the string side thinking - $26$ components $J^{\mu}(I)$. 

This description in terms of ``objects'' is \underline{the main step} in connecting a string system to \underline{our} string field theory formalism.

However, we shall not take all the $J^{\mu}(I)$ variables as just independent variables. Rather we shall represent the ``objects'' needed from string theory by a lower number of degrees of freedom.

\subsection{Development of Reducing ``Object'' - Degrees of Freedom}\label{subsection 5.a}

There are two steps in our representation of the ``objects'' by a lower number of degrees of freedom.

A) We should notice that the freedom of reparametrization in string description has not been totally fixed by the conformal gauge used in section \ref{section 3}. We can indeed further make new ${\hat{\tau}}_{L}(\tau_{L})$ and ${\hat{\tau}}_{R}(\tau_{R})$ arbitrary increasing functions. By such a reparametrization in $\tau_{R}$ and $\tau_{L}$, or in the common variable in the open string case, we can choose one component of $J^{\mu}(I)$ arbitrarily. In the present article we shall choose the infinite momentum frame formalism. I.e. we use the metric (\ref{metric}) and then we can use the reparametrization freedom to ``fix the gauge'' to make (\ref{gf})
\begin{eqnarray}
J^{+}(I)=\frac{a \alpha'}{2}.
\end{eqnarray}

Since the constraints equations (\ref{constraint}) $ \bigl( \dot{X}^{\mu}_{R}(\tau_{R} \bigr) ^{2} = 0 = \bigl( \dot{X}^{\mu}_{L}(\tau_{L} \bigr) ^{2} $ mean that $\dot{X}^{\mu}_{R}(\tau_{R})$ or $\dot{X}^{\mu}_{L}(\tau_{L})$ or $\dot{X}^{\mu}_{R \mathrm{or} L}(\tau)$ in open string case must be light-like, we can really write one of the components of the $J^{\mu}(I)$'s as function of the other ones by using, that the $J^{\mu}(I)$'s are just discretization of the $\dot{X}^{\mu}_{R \mathrm{or} L}(\tau)$, we must have in the approximation of dense discretization
\begin{eqnarray}
\bigl( J^{\mu}(I) \bigr)^{2}=0.
\end{eqnarray}

Then the light-like condition $\bigl( J^{\mu}(I) \bigr) ^{2} =0$ or (\ref{ll}) leads to (\ref{jm}).

So there remain only $25+1-2=24$ independent $J^{\mu}(I)$ components. We call these $24$ remaining the ``transverse '' degrees of freedom.

\subsection{The Even Odd Story \label{subsection 5.b}}

We are forced to make a similar reduction of the degrees of freedom by declaring that only say the $J^{\mu}(I)$ variables corresponding to the ``object'' enumeration number $I$ being even are to be taken as fundamentally existing ``objects'' while the oddly numbered ones should rather be built up from the conjugate variables $\Pi^{i}(I)$ to the $J^{i}(I)$ with even $I$. Indeed we shall take the oddly numbered ones to be built up from the formula (\ref{odd}). We are indeed forced to make a construction of the odd-$I$ object $J^{i}(I)$ in order to make the commutation rules of the match with the commutation rule - say for the right moving fields
\begin{eqnarray}
\bigl[ \dot{X}^{i}_{R}(\tau_{R}), \dot{X}^{k}_{R}(\tau_{R} ') \bigr]
= i2\pi \alpha' \delta^{ik} \frac{\partial}{\partial \tau_{R}} \delta(\tau_{R}'-\tau_{R})
\end{eqnarray}

In fact this kind of derivative of delta- function commutator of the $\dot{X}^{i}_{R}(\tau_{R})$ or $\dot{X}^{i}_{L}(\tau_{R})$ or for open string just $\dot{X}^{i}_{R \mathrm{or} L}(\tau)$ with themselves suggests by the discretization that the neighboring ``objects'' in a cyclically ordered chain of objects must be 
\begin{eqnarray}
\bigl[ J^{i}(I+1), J^{k}(I) \bigr]
&=&\int^{\tau_{R}(I+1+\frac{1}{2})}_{\tau_{R}(I+1-\frac{1}{2})} d\tau_{R}
\int^{\tau_{R}(I+\frac{1}{2})}_{\tau_{R}(I-\frac{1}{2})} d\tau '_{R}
\bigl[ \dot{X}^{i}_{R}(\tau_{R}), \dot{X}^{k}_{R}(\tau_{R} ') \bigr] \\ 
&=&\int^{\tau_{R}(I+1+\frac{1}{2})}_{\tau_{R}(I+1-\frac{1}{2})} d\tau_{R}
\int^{\tau_{R}(I+\frac{1}{2})}_{\tau_{R}(I-\frac{1}{2})} d\tau '_{R} 
\cdot i2\pi \alpha' \delta^{ik} \frac{\partial}{\partial \tau_{R}} \delta(\tau_{R}'-\tau_{R})
\label{com}
\end{eqnarray}
formally.

\begin{figure}[!htb]
\begin{center}
  \includegraphics[width=0.5\textwidth]{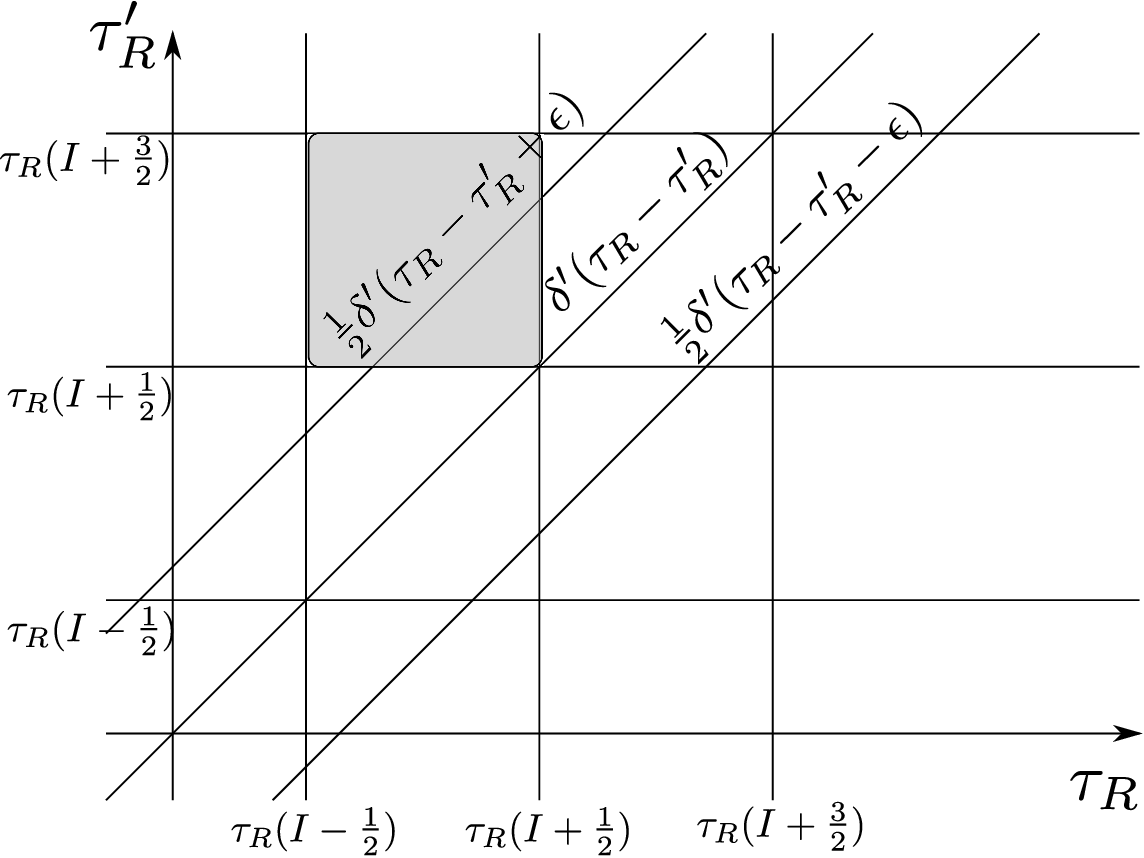}%
\end{center}
\caption{Evaluation of the commutator of two neighboring ``objects'' (an even and an odd) as an integral over the shaded square.  The delta-prime function at first happens to pass just through the border corner so that it is not well defined. We therefore replace it by two by $\epsilon$ displaced delta-prime functions each then with a weight $\frac{1}{2}$. This is also correlated with that we want numerically the same commutator for $J^{i}(I-1)$ and $J^{i}(I)$. \label{figure 8}}
\end{figure}

Now, however, the double integral (\ref{com}) has the derivative of the $\delta$-function just sitting on a kink point of the weight-function and therefore it is not well defined, but we can replace the $\delta'$-function by two half weighted ones
\begin{eqnarray}
\frac{\partial}{\partial \tau_{R}} \delta(\tau_{R}-\tau'_{R})
\rightarrow \frac{1}{2}\frac{\partial}{\partial \tau_{R}} \delta(\tau_{R}-\tau'_{R}-\epsilon)+
\frac{1}{2}\frac{\partial}{\partial \tau_{R}} \delta(\tau_{R}-\tau'_{R}+\epsilon)
\end{eqnarray}
so that we would obtain analogous results for $ \bigl[ J^{i}(I+1), J^{k}(I) \bigr]$ as for $\bigl[ J^{i}(I), J^{k}(I+1) \bigr]$.

Looking e.g. the Figure  \ref{figure 8} we then find that after integrating out for $\tau'_{R}-\tau_{R}>0$ fixed but taking $\tau'_{R}-\tau_{R}<a$ say $\tau'_{R}$ we get
\begin{eqnarray}
\bigl[ J^{i}(I+1), J^{k}(I) \bigr] 
&=&\int^{a}_{0}d(\tau'_{R}- \tau_{R})\cdot (\tau'_{R}- \tau_{R}) \cdot
i2\pi \alpha ' \delta ^{ik} \nonumber \\
&& \qquad \times
  \left( - \frac{\partial}{\partial (\tau'_{R}-\tau_{R})} \delta (\tau'_{R}-\tau_{R}-\epsilon) \right)
  \cdot  \left( \frac{1}{2} \right) \nonumber \\ 
&=&i\frac{2\pi \alpha'}{2} \delta^{ik}= i \pi \alpha' \delta^{ik}
\label{com2}
\end{eqnarray}

Since we denoted the conjugate of $J^{i}(I)$ by $\Pi^{i}(I)$ we have 
\begin{eqnarray}
\bigl[ J^{i}(I), \Pi^{k}(I) \bigr] = i \delta^{ik}
\label{comBl77}
\end{eqnarray}

Then we can arrange the correct commutator (\ref{com2}) by putting the odd numbered $J^{i}(I)$ expressed as a difference in the conjugate momenta $\Pi^{i}(I \pm 1)$ of the neighboring even ``objects''.
\begin{eqnarray}
J^{i}(I)=- \pi \alpha' \left( \Pi^{i}(I+1)-\Pi^{i}(I-1)\right).
\end{eqnarray}

This means that the odd objects can be considered just constructed quantities according to this formula from the conjugate of the even ``objects''. So only the ``transverse'' i.e. the $24$ ``object''-variables $J^{i}(I)$ and their conjugate $\Pi^{i}(I)$, of the \underline{even objects} have to be considered the ``fundamental'' degrees of freedom of our string field theory. The odd objects and the two remaining dimensions are just to be constructed from the $24$ components of only the even ``objects''.

\section{Mass spectrum \label{section 6}}

In single string theory it is well known that the total $25+1$ momentum vector $P^{\mu}$ for, a say, open string is given as 
\begin{eqnarray}
P^{\mu}_{\mathrm{string}} 
&=&\int \frac{\dot{X}^{\mu}(\sigma, \tau)}{2\pi\alpha'}d\sigma \nonumber \\ 
&=&\int \frac{\dot{X}^{\mu}_{R}(\tau-\sigma)+\dot{X}^{\mu}_{L}(\tau+\sigma)}{2\pi\alpha'}d\sigma \nonumber \\ 
&=&\frac{1}{2\pi\alpha'}\left( \sum J^{\mu}_{R}(I) + \sum J^{\mu}_{L}(I) \right) \nonumber \\
&=&\frac{1}{2\pi\alpha'}\sum _{I=0}^{N-1} J^{\mu}(I)
\end{eqnarray}
where we have taken the ``object'' enumeration number $I$ going in the case of open string over both right and left movers - which so to speak mix up for open string - to run around the circular ordering $I=0,1,2,3, \dots, N-1$. 

Now we shall use this formula for the ($25+1$)-momentum vector for a string, which in turn corresponds to one of our cyclically ordered chains of ``objects'', of which only half are the ``fundamental even ones'', while the other half are the constructed odd ones. Because of this even-odd-distinction the total number $N$ of ``objects'' in the chain must be even.

We can for instance ask for the mass square 
\begin{eqnarray}
M^{2}&=&2P^{+}P^{-}-\sum^{24}_{i=1}(P^{i})^{2}  \nonumber \\
&=&2 \left( \sum^{N-1}_{I=0}J^{+}(I) \right) \left( \sum^{N-1}_{I=0}J^{-}(I) \right) 
\left( \frac{1}{2\pi\alpha'} \right)^{2} \nonumber \\ 
&& -\sum^{24}_{i=1}\left( \sum^{N-1}_{I=0}J^{i}(I) \right)^{2} 
 \frac{1}{(2\pi\alpha')^{2}}. 
 \label{M2}
\end{eqnarray}
This formula at the end can be written as an operator acting on the quantum state of the $\frac{N}{2}$ ``fundamental even objects'' that are described by their ``fundamental''meaning ``transverse'' degrees of freedom. This is to say that the mass square operator is an operator acting on wave functions $\psi$ depending on $\frac{N}{2} \cdot 24 J^{I} (I)$ coordinates with $I=0,2,4, \dots, N-4, N-2$ and $i=1,2, \dots, 23,24$.

In this section we want to show that the spectrum of the mass square $M^{2}$ is (apart from the zero point energy) identical to that of an open string just by using the constructions of other ``objects'' and their components mentioned above.

To achieve this derivation of the mass square spectrum we shall first simplify a bit some of the expressions occurring in (\ref{M2}). In fact we easily see 
\begin{eqnarray}
\sum^{N-1}_{I=0}J^{+}(I)=\frac{a \alpha'}{2}\cdot N
\end{eqnarray}
and
\begin{eqnarray}
\sum^{N-1}_{I=0}J^{-}(I)&=&\sum^{N-2}_{I=0 \atop I \ \mathrm{even}} \ J^{-}(I)
+\sum^{N-1}_{I=1 \atop I \ \mathrm{odd}} \ J^{-}(I)\nonumber \\ 
&=&\frac{1}{a \alpha'}\sum^{N-2}_{I=0 \atop I \ \mathrm{even}} \sum^{24}_{i=1}\left( J^{i}(I)\right)^{2}
+\frac{\pi^{2}\alpha'}{a}\sum^{N-1}_{I=1 \atop I \ \mathrm{odd}} 
\sum^{24}_{i=1}\left( \Pi^{i}(I+1)-\Pi^{i}(I-1) \right)^{2}.
\end{eqnarray}

The term inside the bracket of the last term (\ref{M2}) reads
\begin{eqnarray}
\sum^{N-1}_{I=0}J^{i}(I)&=&\sum^{N-2}_{I=0 \atop I \ \mathrm{even}}J^{i}(I)
+\sum^{N-1}_{I=1 \atop I \ \mathrm{odd}}J^{i}(I) \nonumber \\ 
&=&\sum^{N-2}_{I=0 \atop I \ \mathrm{even}}J^{i}(I)
+\sum^{N-1}_{I=1 \atop I \ \mathrm{odd}}(-\pi\alpha')\left( \Pi^{i}(I+1)-\Pi^{i}(I-1)\right).
 \label{6.5}
\end{eqnarray}

So the first term in (\ref{M2}) reads
\begin{eqnarray}
\frac{2}{(2\pi\alpha')^{2}}\left( \sum^{N-1}_{I=0}J^{+}(I) \right) \left( \sum^{N-1}_{I=0}J^{-}(I) \right) 
&=&\frac{2}{(2\pi \alpha')^{2}}\frac{a \alpha'}{2}\cdot N \cdot \biggl( \frac{1}{a \alpha'}\sum^{N-2}_{I=0 \atop I \ \mathrm{even}}\sum^{24}_{i=1}\left( J^{i}(I)\right)^{2} \nonumber \\
&&+\frac{\pi^{2}\alpha'}{a}\sum^{N-1}_{I=1 \atop I \ \mathrm{odd}}\sum^{24}_{i=1} 
\left( \Pi^{i}(I+1)-\Pi^{i}(I-1) \right)^{2} \biggr) \nonumber \\ 
&=&\frac{N}{(2\pi\alpha')^{2}}\sum^{N-2}_{I=0 \atop I \ \mathrm{even}}\sum^{24}_{i=1} \left( J^{i}(I) \right)^{2} 
\nonumber \\
&&+\frac{N}{4}\sum^{N-1}_{I=1 \atop I \ \mathrm{odd}}\sum^{24}_{i=1} 
\left( \Pi^{i}(I+1)-\Pi^{i}(I-1) \right)^{2}
\label{6.6}
\end{eqnarray}

Like the last term (\ref{6.5}) this term (\ref{6.6}) is also bilinear and the ``fundamental'' variables $\Pi^{i}(I)$, $J^{i}(I)$ for $I$ even and $i=1,2, \dots,24$ and thus the whole expression for the mass square operator is 
\begin{eqnarray}
M^{2}=\frac{1}{(2\pi\alpha')^{2}} \sum^{24}_{i=1}\left( N \sum^{N-2}_{I=0 \atop I \mathrm{even}}\left(J^{i}(I) \right)^{2}
-\sum^{N-2}_{\stackrel{K=0}{K \ \mathrm{even}}}\sum^{N-2}_{\stackrel{I=0} {I \ \mathrm{even}}}J^{i}(I)J^{i}(K) \right) \nonumber \\ 
+\frac{N}{4}\sum^{24}_{i=1}\sum^{N-1}_{I=1 \atop I \ \mathrm{odd}}\left( \Pi^{i}(I+1)-\Pi^{i}(I-1) \right)^{2}
\label{6.7}
\end{eqnarray}

Since this expression is purely quadratic it can be resolved into $N$ harmonic oscillators - except that one of them corresponding to a free particle of which position is the average of all the $\Pi^{i}(I)$'s - and e.g. its spectrum is exactly calculable.
 
To perform this calculation of the spectrum (\ref{6.7}) by resolving the problem into the $N$(or $N-1$) harmonic oscillators we shall make use of the $Z_{\frac{N}{2}}$ symmetry group of pushing all the even ``objects'' around in the cyclically ordered chain. Using this symmetry means that we use a Fourier series expansion by writing the ``fundamental'' $J^{i}(I)$, $\Pi^{i}(I)$ for even $I$ ($I=0,2,4,\dots,N-4,N-2$) and $(i=1,2, \cdots, 24)$ as 
\begin{eqnarray}
J^{i}(I)&=&\mathrm{Re}\left( \sum^{N-1}_{L=0} c^{i}_{L}\mathrm{exp} \left(\frac{iL \cdot I 2\pi}{N} \right) \right)\nonumber \\ 
&=&2\mathrm{Re}\left( \sum^{\frac{N}{2}-1}_{L=0} c^{i}_{L}\mathrm{exp} \left(\frac{iL \cdot I 2\pi}{N}\right) \right) \nonumber \\
&&\mathrm{for \ even} \ I \ \mathrm{and} \ i=1,2,\dots,24.
\label{repcBl93} 
\end{eqnarray}
and
\begin{eqnarray}
\Pi^{i}(I)&=&\mathrm{Re}\left( \sum^{N-1}_{L=0} d^{i}_{L}\mathrm{exp} \left(\frac{iL \cdot I 2\pi}{N}\right) \right) \nonumber \\ 
&=&2\mathrm{Re}\left( \sum^{\frac{N}{2}-1}_{L=0} d^{i}_{L}\mathrm{exp} \left(\frac{iL \cdot I 2\pi}{N}\right) \right).
\label{repdBl93}
\end{eqnarray}
However, the $c^{i}_{L}$'s and $d^{i}_{L}$'s are not independent, but rather obey
\begin{eqnarray}
c^{i}_{L}=c^{i}_{L+\frac{N}{2}}=(c^{i}_{-L})
\end{eqnarray}
and analogously
\begin{eqnarray}
d^{i}_{L}=d^{i}_{L+\frac{N}{2}}=(d^{i}_{-L})
\end{eqnarray}
so that in reality only $\frac{N}{4}$ out of the at first $N$ complex number are independent. Thus we should at the end only sum over $\frac{N}{4}$ different $L$-values, and we get relations
\begin{eqnarray}
J^{i}(I)&=&\left[1+(-1)^{I}\right] \sum^{\frac{N}{4}-1}_{L=0}2(\mathrm{Re} \ c^{i}_{L})
\cdot \cos \frac{L \cdot I \cdot 2\pi}{N}
\nonumber \\
&&+ \left[1+(-1)^{I}\right] 
\cdot \sum^{\frac{N}{4}-1}_{L=1}2 (\mathrm{Im} \ c^{i}_{L}) \sin \frac{L \cdot I \cdot 2\pi}{N}\nonumber \\ 
&=&4\sum^{\frac{N}{4}-1}_{L=0}\mathrm{Re} \ (c^{i}_{L})\cos \frac{L \cdot I \cdot 2\pi}{N}
+4 \sum^{\frac{N}{4}-1}_{L=0} \mathrm{Im} \ c^{i}_{L} \cdot \sin \frac{L \cdot I \cdot 2\pi}{N}\nonumber \\ 
&&(\mathrm{for \ even} \ I)
\end{eqnarray}
and analogously for the conjugate momenta
\begin{eqnarray}
\Pi^{i}(I)&=&\left[ 1+(-1)^{I} \right] \cdot \sum^{\frac{N}{4}-1}_{L=0}\left[ (\mathrm{Re} \ d^{i}_{L})\cos \frac{2\pi L \cdot I}{N}
+(\mathrm{Im} \ d^{i}_{L})\cdot \sin \frac{2\pi L \cdot I}{N} \right]\nonumber \\ 
&=&4\sum^{\frac{N}{4}-1}_{L=0}2(\mathrm{Re} \ d^{i}_{L})\cos \frac{2\pi L \cdot I}{N}
+4 \sum^{\frac{N}{4}-1}_{L=1} (\mathrm{Im} \ d^{i}_{L}) \sin \frac{2\pi L \cdot I}{N}.\nonumber \\
&&(\mathrm{for \ even} \ I)
\end{eqnarray}
For odd values of $I$ we have arranged to get zero since the factor $1+(-1)^{I}$ suggests by our conditions
\begin{eqnarray}
c^{i}_{L}=c^{i}_{L+\frac{N}{2}} \ \mathrm{and} \ d^{i}_{L}=d^{i}_{L+\frac{N}{2}}. 
\end{eqnarray}
These Fourier series expansions can essentially as usually be inverted to
\begin{eqnarray}
c^{i}_{L}=\frac{1}{N}\sum^{N-2}_{I=0,2,4,\dots \atop I \ \mathrm{even}}J^{i}(I)e^{-\frac{i2\pi L \cdot I}{N}}
\label{repBl97}
\end{eqnarray}
and
\begin{eqnarray}
d^{i}_{L}=\frac{1}{N}\sum^{N-2}_{I=0,2,4,\dots \atop I \ \mathrm{even}}\Pi^{i}(I)e^{-\frac{i2\pi L \cdot I}{N}}
\label{repdBl97}
\end{eqnarray}
and so we derive from the commutation relation (\ref{comBl77})
\begin{eqnarray}
\left[ J^{i}(I), \Pi^{k}(I) \right] = i \delta^{ik}
\end{eqnarray}
so that
\begin{eqnarray}
[c^{i}_{L},(d^{k}_{L})^{*}]=\frac{i}{2N}\delta_{LK} \delta^{ik}
\end{eqnarray}
and
\begin{eqnarray}
[c^{i}_{L},(d^{k}_{L})]=\frac{i}{2N}\delta_{K,-L} \delta^{ik}
\label{comBl98}
\end{eqnarray}
and thus
\begin{eqnarray}
[\mathrm{Re} \ c^{i}_{L},\mathrm{Re} \ d^{k}_{K})]=\frac{i}{4N}\delta_{LK} \delta^{ik}
\end{eqnarray}
\begin{eqnarray}
[\mathrm{Im} \ c^{i}_{L},\mathrm{Im} \ d^{k}_{K})]=\frac{i}{4N}\delta_{KL} \delta^{ik}.
\label{comBl98.t}
\end{eqnarray}

In order to express our total $P^{-}_{\mathrm{tot}}$-component in terms of these $c^{i}_{L}$ and $d^{i}_{L}$ variables, we first express $P^{-}_{\mathrm{tot}}$ by the cyclically ordered chain of $\frac{N}{2}$ even and $\frac{N}{2}$ odd ``objects'',
\begin{eqnarray}
M^{2}=2P^{+}_{\mathrm{tot}}P^{-}_{\mathrm{tot}}-\sum^{24}_{i=1}(P^{i}_{\mathrm{tot}})^{2}
\end{eqnarray}
is given as
\begin{eqnarray}
P^{-}_{\mathrm{tot}}&=&P^{-}_{\mathrm{tot}}|_{\mathrm{even \ part}} + P_{\mathrm{tot}}|_{\mathrm{odd \ part}} \nonumber \\ 
&=&\frac{1}{2\pi\alpha'}(J^{-}_{\mathrm{tot}}|_{\mathrm{even \ part}} + J^{-}_{\mathrm{tot}}|_{\mathrm{odd \ part}} )
\end{eqnarray}
where then
\begin{eqnarray}
P^{-}_{\mathrm{tot}}|_{\mathrm{even \ part}} &=& \frac{1}{2\pi\alpha'}(J^{-}_{\mathrm{tot}}|_{\mathrm{even \ part}}\nonumber \\ 
&=&\frac{N}{4\pi(\alpha')^{2}a}\sum^{24}_{i=1}\sum^{\frac{N}{2}-1}_{L=0}(c^{i}_{L})^{*}\cdot c^{i}_{L} \\ \nonumber
&=&\frac{N \cdot 2}{4\pi(\alpha')^{2}a}\sum^{24}_{i=1}\sum^{\frac{N}{4}-1}_{L=0}|c^{i}_{L}|^{2}
\end{eqnarray}
and
\begin{eqnarray}
P^{-}_{\mathrm{tot}}|_{\mathrm{odd \ part}} &=& \frac{1}{2\pi\alpha'}(J^{-}_{\mathrm{tot}}|_{\mathrm{odd \ part}}\nonumber \\ 
&=&\frac{4\pi N}{\alpha}\sum^{24}_{i=1}\sum^{\frac{N}{2}-1}_{L=0}|d^{i}_{L}|^{2}
{\sin}^{2} \frac{2 \pi L}{N}.
\end{eqnarray}
Using that
\begin{eqnarray}
P^{+}_{\mathrm{tot}}=\frac{1}{2 \pi \alpha'}J^{+}_{\mathrm{tot}}=\frac{1}{2 \pi \alpha'}\cdot N \cdot \frac{a \alpha'}{2}
=\frac{aN}{4\pi},
\end{eqnarray}
we get the expression
\begin{eqnarray}
M^{2}&=&\frac{N^{2}}{8 \pi^{2}(\alpha')^{2}} \sum^{24}_{i=1}\sum^{\frac{N}{2}-1}_{L=0}|c^{i}_{L}|^{2}
+2N^{2}\sum^{24}_{i=1}\sum^{\frac{N}{2}-1}_{L=0}|d^{i}_{L}|^{2} \sin^{2}\frac{2\pi L}{N}
-\sum^{24}_{i=1}(P^{i}_{\mathrm{tot}})^{2} \nonumber \\ 
&=&\frac{N^{2}}{4\pi^{2}(\alpha')^{2}}\sum^{24}_{i=1}\sum^{\frac{N}{4}-1}_{L=0}|c^{i}_{L}|^{2}
+4N^{2}\sum^{24}_{i=1}\sum^{\frac{N}{4}-1}_{L=0}|d^{i}_{L}|^{2} \sin^{2}\frac{2\pi L}{N}.
\label{M2f}
\end{eqnarray}

From here we recognize that $M^{2}$ appears written as the sum of two series of harmonic oscillator Hamiltonians. One of them is rewritten in terms of $q$-coordinate Re $d^{i}_{L}$ with the momentum being Re $c^{i}_{i}$, (the other one comes from the imaginary part, see below) or to make the usual commutation out of (\ref{comBl98.t}) take
\begin{eqnarray}
p^{i}_{L(r)}\hat{=}-4N \ \mathrm{Re} \ c^{i}_{L}
\end{eqnarray}
so that
\begin{eqnarray}
[\mathrm{Re \ d^{i}_{L}}, p^{k}_{K}]=i \ \delta^{ik}\delta_{LK}.
\end{eqnarray}
The Hamiltonian of the harmonic oscillators of which occurs in the mass square operator $M^{2}$ from (\ref{M2f}) are then one series
\begin{eqnarray}
H^{i}_{L(r)}=\frac{(p^{i}_{L(r)})^{2}}{2m^{i}_{r}}+\frac{1}{2}m^{i}_{L(r)}(\omega^{i}_{L(r)})^{2}(\mathrm{Re} \ d^{i}_{L})^{2}
\end{eqnarray}
where then
\begin{eqnarray}
\frac{1}{2m^{i}_{L(r)}}=\frac{2}{128 \pi^{2}(\alpha')^{2}}
\end{eqnarray}
and
\begin{eqnarray}
\frac{1}{2}m^{i}_{L(r)}(\omega^{i}_{L(r)})^{2}=4N^{2}\sin^{2} \frac{2 \pi L}{N}.
\end{eqnarray}
This means that the frequency of the oscillators in this series when the $M^{2}$ is considered the Hamiltonian (or energy) is
\begin{eqnarray}
\omega^{i}_{L(r)}&=&\sqrt{4 \cdot \frac{2}{128 \pi^{2}(\alpha')^{2}}\cdot 4N^{2}\sin^{2}\frac{2 \pi L}{N}}\nonumber \\ 
&=&\frac{1}{2\pi \alpha'} \cdot N \sin \frac{2\pi L}{N}
\label{olrBl104}
\end{eqnarray}
which for $L<<N$ reduces to 
\begin{eqnarray}
\omega^{i}_{L(r)}=\frac{1}{2\pi \alpha'} \cdot 2\pi L =\frac{1}{\alpha'}\cdot L\ \ \ \ \ \mathrm{for} \ L<<N.
\end{eqnarray}

Similarly the series of imaginary parts are considered delivering to $M^{2}$ harmonic oscillator Hamiltonian terms
\begin{eqnarray}
H^{i}_{L(\mathrm{im})}=\frac{(p^{i}_{L(\mathrm{im})})^{2}}{2m^{i}_{L(\mathrm{im})}}+\frac{1}{2}(\mathrm{Im} \ d^{i}_{L})^{2}m^{i}_{L(\mathrm{im})}
\cdot (\omega^{i}_{L(\mathrm{im})})^{2}
\end{eqnarray}
where then 
\begin{eqnarray}
p^{i}_{L(\mathrm{im})}&=&-4N \ \mathrm{Im} \ c^{i}_{L}\mathrm{and}\nonumber \\ 
\frac{1}{2m^{i}_{L(\mathrm{im})}}&=&\frac{N^{2}\cdot 2}{8 \pi^{2}(\alpha')^{2}}\cdot\frac{1}{(4N)^{2}} \nonumber \\ 
&=&\frac{1}{64 \pi^{2}(\alpha')^{2}}
\end{eqnarray}
and
\begin{eqnarray}
\frac{1}{2}m^{i}_{L(\mathrm{im})}(\omega^{i}_{L(\mathrm{im})})^{2}=4N^{2}\sin^{2}\frac{2\pi L}{N}.
\end{eqnarray}
So
\begin{eqnarray}
(\omega^{i}_{L(\mathrm{im})})^{2}&=&4 \cdot 4N^{2} \sin^{2}\frac{2 \pi L}{N}\cdot \frac{1}{64\pi^{2}(\alpha')} \nonumber \\ 
&=&\frac{1}{(2\pi \alpha')^{2}} \cdot N^{2}\sin^{2}\frac{2\pi L}{N}.
\label{olimBl106}
\end{eqnarray}
Taking the limit $L<<N$ this means
\begin{eqnarray}
\omega^{i}_{L(\mathrm{im})}=\frac{2\pi L}{2\pi \alpha'}=\frac{L}{\alpha'} \ .
\label{olimBl107}
\end{eqnarray}

Since $L$ runs through the integers from $0$ or $1$ up to $\frac{N}{4}$ we see that in the $L<<N$ region these frequencies just give the spectrum of the dual string models except that we got two series.

\subsection{Species Doubler Problem \label{subsection 6-A}}

When we Fourier resolved the $J^{i}(I)$ and $\Pi^{i}(I)$ into a Fourier series with coefficients $c^{i}_{L}$ and $d^{i}_{L}$ respectively we resolved the two fields separately.

When we at the end made the long wave length approximation by taking $L<<N$ it were the hope to peak up physically a $J^{i}(I)$-series varied very smoothly or slowly with the integer variable $I$. As we did our Fourier resolution using even $I$ only - because they were considered ``fundamental'' in our model - it is not guaranteed that we shall get smoothness as function of $I$, if we also include the odd values of $I$ just by taking the $L$ small.

Rather we should explicitly investigate for which $L<<N$ values of $L$ and which series of eigenvalues we do indeed have an even $I$ with odd values included smooth $J^{i}(I)$.

Accordingly to our used definition of $J^{i}(I)$ for odd $I$, namely (\ref{odd}) a smooth $J^{i}(I)$ as function of $I$ for both $I$ even and odd must mean that we approximately must have from continuity for $I$ odd
\begin{eqnarray}
J^{i}(I-1)\approx -\pi \alpha'\left( \Pi^{i}(I+1)-\Pi^{i}(I-1) \right)\approx J^{i}(I+1).
\end{eqnarray}
Now we insert herein Fourier representations (\ref{repdBl97}) and (\ref{repcBl93}) or rather (\ref{repcBl93}) and (\ref{repdBl93}) to give us the smoothness requirement
\begin{eqnarray}
&&2\mathrm{Re}\left( \sum^{\frac{N}{2}-1}_{L=0} c^{i}_{L}\mathrm{exp}\frac{i L (I\pm 1)2\pi}{N} \right)  \nonumber \\
&& \qquad \approx-\pi \alpha' 2\mathrm{Re}\Biggl( \sum^{\frac{N}{2}-1}_{L=0} d^{i}_{L} \mathrm{exp} \left( \frac{i L (I+1)2\pi}{N}\right)
-\mathrm{exp}\left( \frac{i L (I-1)2\pi}{N} \right) \Biggr) \end{eqnarray}
which for $L<<N$ leads to
\begin{eqnarray}
c^{i}_{L}\approx -i \pi\alpha' \cdot 2 \cdot 2\pi\frac{1}{N}d^{i}_{L}\cdot L
=-i 4\pi^{2} \cdot \frac{\alpha'}{N} \cdot d^{i}_{L}\cdot L.
\label{smoothtBl5}
\end{eqnarray}

Now let us remember that we have at first for a given value of $L$ due to the two series of solutions $(\mathrm{r})$ and $(\mathrm{im})$ \underline{two} eigenmodes of the same frequency 
\begin{eqnarray}
\omega^{i}_{L(\mathrm{r})}=\omega^{i}_{L(\mathrm{im})}
\end{eqnarray}
according to (\ref{olrBl104}) and (\ref{olimBl106}). If they were only one of these two modes involving respectively for $(\mathrm{r})$ and $(\mathrm{im})$ only the real or imaginary parts of $c^{i}_{L}$ and $d^{i}_{L}$ the smoothness condition (\ref{smoothtBl5}) would \underline{not} be fulfilled. Only by exciting the two vibration modes denoted $(L,i,(\mathrm{r}))$ and $(L,i,(\mathrm{im}))$ respectively in a correlated way can we obtain the smoothness condition (\ref{smoothtBl5}). For seeing this the explicit $i$ in equation (\ref{smoothtBl5}) is crucial. These facts mean that we shall count half the modes of vibrations of string or rather cyclically ordered chain in our model as (a kind of) ``species doubler modes''. That is to say it is only half of the modes found at first with small $L$ which actually vibrate in an even for odd $I$'s smooth way. Thus we should only think of those modes which are truly smooth or continuous in the sense of obeying (\ref{smoothtBl5}) as co
 ntinuous string-vibration modes. This interpretation will bring the spectrum of our cyclically ordered chains to agree \underline{exactly} with the open string spectrum.

But we must seek to understand the physical reason for these unwanted doubling of the modes of vibration of the string which we obtained at first by using ``object''-model:

It is obvious that when we have in our model being identified with right and left mover modes a discretization it is analogous to having introduced a lattice in the $\tau_{R}$ and $\tau_{L}$ coordinates and then one cannot have only right-mover field, but will always obtain also species doublers as is well known in lattice field theory \cite{19}. So our smoothness condition (\ref{smoothtBl5}) is destined to forbid by assumption so to speak the unavoidable species doubler in a lattice. But if we accept such a smoothness condition then we got the spectrum to be perfectly agreeing to that of string theory. The formalism put forward in the present article of course corresponds to the light-cone gauge formulation of string theory, so that according to Goddard, Goldstone, Rebbi and Thorn \cite{20} the theory is in danger of not being Lorentz invariant but manages to be so for the very special dimension $25+1$ for our bosonic theory.

\section{Veneziano Model \label{section 7}} 

In this article we shall not really deliver the full derivation of the scattering amplitudes for strings in our ``objects''-based string field theory, but rather postpone it to a later article \cite{22} since it takes quite a bit of algebra.

Let us, however, review the philosophy of scattering in our model, namely that we do not consider scattering ontologically taking place, but rather it means an only formal - not physical - reshuffling of the cyclically ordered chains of ``objects'' into a new combination: or classification into cyclically ordered chains. A bit provocatively states: Nothing happens when the strings scatter on each other. It is only that some series of even ``objects'' get split up and classified into cyclically ordered chains in a new way.

But all these even ``objects'' have the same $J^{i}(I)$ and $\Pi^{i}(I)$, ($I$ even) before and after the unification or splitting of the chains. Considering the odd-``objects'', however, there is a tiny change, because each ``odd object'' is constructed from its two \underline{neighboring} even ones, namely the $J^{i}(I)$ for odd $I$ were proportional to the difference of the conjugate momenta $\Pi^{i}(I+1)$ and $\Pi^{i}(I-1)$ for the neighboring even ``objects''. But now the point is that, if one reshuffles the way the even ``objects'' are put into cyclically ordered chains, then there will disappear and appear pairs of next to neighbor even ``objects'' and thus some previously included ``odd objects'' will disappear while others will appear, to replace them, so to speak. This process is only justified by continuity assumption the states of which has to be discussed more later. We shall consider as the simplest case an open string splitting into two. That is to say that one
  cyclically ordered chain of event which is associated with construction of the ``odd objects'', splits into two cyclically ordered chains that corresponds to the successive pairs of next neighbors.

\begin{figure}[!htb]
\begin{center}
\includegraphics[width=0.5\textwidth]{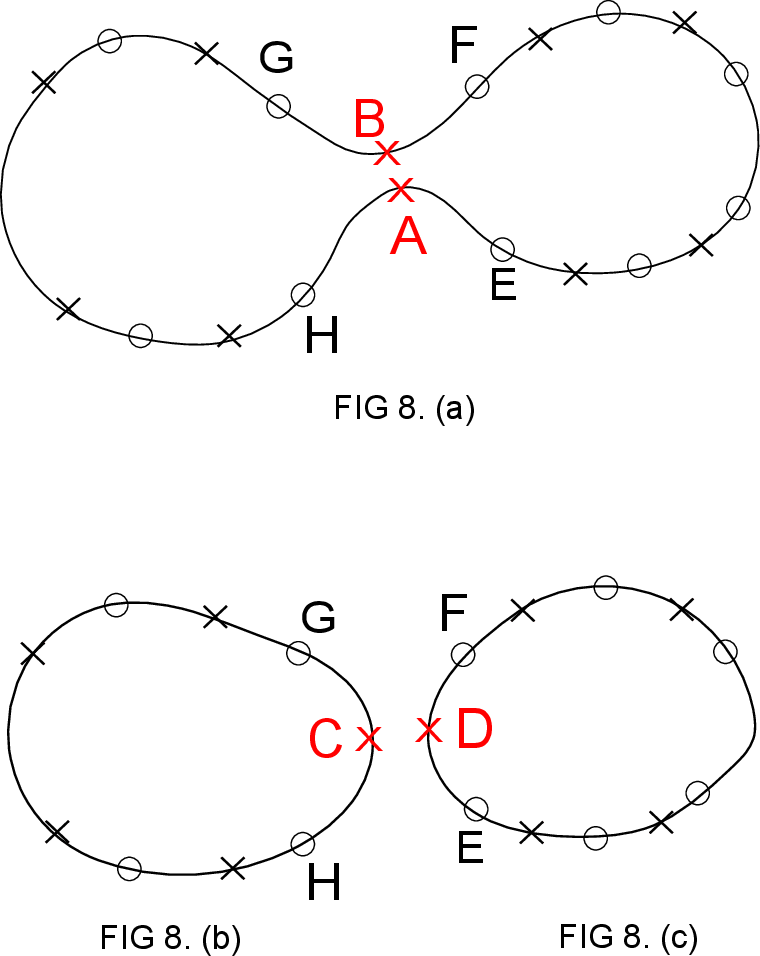}%
\end{center}
\caption{We illustrate cyclically ordered chain in Figure  \ref{figure 9}(a), which formally splits into two separate chains depicted in Figure  \ref{figure 9}(b) and (c). In the Figure  \ref{figure 9} all the even ``objects'' are denoted as small circles $ \circ$ while all the odd ``objects'' we denote as crosses $ \times $. In the original cyclically ordered chain, Figure  \ref{figure 9}(a) two of the odd objects denoted by red colored A and B are removed by splitting. Then to make up two into new cyclically ordered chains all the even objects such as E, F, G and H refound after the splitting into two chains. Furthermore two \underline{new} odd ``objects'' denoted by red colored C and D are created in each new chains as depicted in Figure  \ref{figure 9}(b) and (c) respectively.\label{figure 9}}
\end{figure}

On the Figure  \ref{figure 9} we show the simplest splitting which consists of two neighboring places in the chain to change the next even neighbor. We illustrate the three cyclically ordered chains: The Figure  \ref{figure 9}(a) shows the cyclically ordered chain before splitting, while the Figure  \ref{figure 9}(b) and (c) are those after splitting into the two chains. Here a remark is in order. Under this simplest splitting of one cyclically ordered chain in Figure  \ref{figure 9}(a) into two Figure  \ref{figure 9}(b) and (c), two of the ``odd objects'' A and B in Figure  \ref{figure 9}(a) are replaced by two new ``odd objects'' C and D in Figure  \ref{figure 9}(b) and (c) respectively. 

If we denote these ``odd objects'' corresponding $J^{\mu}(I)$ as $J^{\mu}(A)$, $J^{\mu}(B)$, $J^{\mu}(C)$ and $J^{\mu}(D)$ and have in mind that the total ($25+1$)-momentum is $\frac{1}{2\pi\alpha'}$ times the sum over all the $J^{\mu}(I)$'s then the $(25+1)$-momentum gets by this scattering or decay reshuffling changed by the amount 
\begin{eqnarray}
J^{\mu}(C)+J^{\mu}(D)-J^{\mu}(A)-J^{\mu}(B)
\end{eqnarray}
Luckily enough it is trivially seen that for the transverse components this change reads
\begin{eqnarray}
&&J^{i}(C)+J^{i}(D)-J^{i}(A)-J^{i}(B) \nonumber \\
&& \quad =-\pi\alpha' \left( \Pi^{i}(H)-\Pi^{i}(G) \right) + \left( \Pi^{i}(F)-\Pi^{i}(E) \right) \nonumber \\
&& \qquad \ - \left( \Pi^{i}(H)-\Pi^{i}(E) \right)  - \left( \Pi^{i}(F)-\Pi^{i}(G) \right) \nonumber \\
&& \quad =0
\end{eqnarray}
so that indeed the ``transverse'' momentum is trivially conserved under such a ``scattering''.  Since the
\begin{eqnarray}
P^{+}=\frac{1}{2\pi\alpha'}J^{+}=\frac{1}{2\pi\alpha'}\sum_{\mathrm{all~ objects}}J^{+}
=\frac{a}{4\pi}\sharp  (\mathrm{objects}),
\end{eqnarray}
the $P^{+}$-component of momentum is also seen to be conserved. However, the question of conservation of the minus component $P^{-}$ gets non-trivial and leads to a need for an extra condition at the splitting point. Inserting the more complicated expressions for $J^{-}(A), J^{-}(B), J^{-}(C)$ and $J^{-}(D)$ from equation 
\begin{eqnarray}
J^{-}(I)&=&\frac{1}{a\alpha'}\cdot(\pi\alpha')\sum^{24}_{i=1}\left( \Pi^{i}(I+1)-\Pi^{i}(I-1)\right)^{2} \nonumber \\ 
&=&\frac{\pi^{2}\alpha'}{a}\sum^{24}_{i=1}\left( \Pi^{i}(I+1)-\Pi^{i}(I-1)\right)^{2} 
\end{eqnarray}
into the expression for the change in total sum over the $J^{-}$'s including the odd ones leads to
\begin{eqnarray}
\lefteqn{J^{-}(C)+J^{-}(D)-J^{-}(A)-J^{-}(B)  } \nonumber  \\
&&=\frac{\pi^{2}\alpha'}{a}\sum^{24}_{i=1}\biggl[ \left( \Pi^{i}(H)-\Pi^{i}(G)\right)^{2}
+\left( \Pi^{i}(F)-\Pi^{i}(E)\right)^{2} \nonumber \\
&&\qquad \qquad \qquad -\left( \Pi^{i}(F)-\Pi^{i}(G)\right)^{2}
-\left( \Pi^{i}(H)-\Pi^{i}(E)\right)^{2} \biggr] \nonumber \\ 
&& =\frac{2\pi^{2}\alpha'}{a}\sum^{24}_{i=1}\biggl[ - \Pi^{i}(H)\Pi^{i}(G)-
\Pi^{i}(F)\Pi^{i}(E)+ \Pi^{i}(F)\Pi^{i}(G)
+\Pi^{i}(H)\Pi^{i}(E)\biggr]  \nonumber \\ 
&&=\frac{2\pi^{2}\alpha'}{a}\sum^{24}_{i=1}\left( - \Pi^{i}(F)-\Pi^{i}(H) \right)
\left(\Pi^{i}(G)-\Pi^{i}(E) \right).
\end{eqnarray}

This expression could be made $0$ provided either
\begin{eqnarray}
\Pi^{i}(G)=\Pi^{i}(E)
\end{eqnarray}
or
\begin{eqnarray}
\Pi^{i}(F)=\Pi^{i}(H).
\end{eqnarray}
The hope should be that such a condition should correspond to the case that the open string to decay by ``hitting itself'' or rather does whatever is needed for decaying.

\subsection{Plan for Scattering Amplitude Calculation{\label{sebsection 7}}}

To truly calculate a scattering or say even simpler a decay amplitude in our model one should consider:

a) The external strings for which one wants the scattering amplitude correspond - if open - just each to a certain cyclically ordered chain of both odd and even objects. One can then look for the wave function $\psi \bigl( \bigl\{ J^{i}(I), I \ \mathrm{even} \bigr\} \bigr)$ in the space of the $\frac{N}{2}\cdot 24 \ J^{i}$-coordinates representing just the selected eigenstate of the mass square to be scattered.  

b) By composing the wave functions for all the incoming string states expressed as states of even objects by wave functions for these (even) objects, we obtain a wave function in $\frac{N_{1}+N_{2}+\cdots+N_{e}}{2}\cdot 24$ (where $N_{1},N_{2}, \cdots $ are the numbers of objects in the various incoming strings) dimensional $J$-component space describing the whole incoming state in terms of even objects.

In an analogous way we can construct a state in terms of the even ``objects'' describing the outgoing state when one has decided to calculate an $S$-matrix element.

c) Now the story that nothing really happens in the scattering - it be all fake - is actually to mean that we formally like to play with an $S$-matrix that is just the unit matrix. This means that we simply consider the overlap
\begin{eqnarray}
<f|i>\hat{=}&&\int \psi^{*}_{\mathrm{outgoing}}\left( J^{i}_{\mathrm{even} \ I}(I) \ \mathrm{for \ all \ outgoing \ particles} \right) \\ \nonumber 
&& \ \ \ \ \psi_{\mathrm{incomming}} \left( J^{i}_{\mathrm{even} \ I}(I) \ \mathrm{for \ all \ incomming \ particles} \right)dJ
\end{eqnarray}

d) Then the idea is that although the overlap or inner product $<f|i>$ calculated under c) is just an overlap formally it is our expectation and hope that it will turn out to be a function of the momenta and internal states used to construct the associated ``even-object'' states. This overlap $<f|i>$ looks (numerically) quite like a scattering amplitude. Really we have the expectation that the overlap $<f|i>$ will be just the $S$-matrix element in the Veneziano model \cite{21}.

e) Really there are very good hopes that this string field theory of ours - which is a rewriting of string theory - should lead to the Veneziano model. Indeed it will be needed to reshuffle the (even) objects from the various cyclically ordered chains in the initial state to obtain a classification into such chains matching the final state. But now we expect that the overlap $<f|i>$ which we calculate will be dominated by those contributions for which most of the even `` objects'' can keep their neighbors which they had in the initial state also in the final state. That is to say that the splitting and reuniting of these cyclically ordered chains should be as few as possible. In that approximation the overlap $<f|i>$ to be identified with the scattering amplitude would appear as a sum over all the possible places for these necessary splittings and unifications being needed. Since the sums are over discrete variables that are at the end to be made continuous when $a \to0$, the
 se sums of course are only approximations to integrals. These integrals run over imagined or faked happenings each run of which has in fact an interpretation in terms of strings. So it looks very much like the integral representations of Veneziano model (on the surface at least).

\section{Discussion and Outlook \label{section 8}}

We have presented a proposal for a new string field theory in which the hang together of the strings in the topological way is not taken to be a truly ontologically existing degrees of freedom. Rather a system of several, say open strings are described by what we call ``objects'': each open string corresponds to a cyclically ordered chain of such objects. These ``objects'' represent a discretization of say $\dot{X}^{\mu}_{R}(\tau-\sigma)=\dot{X}^{\mu}_{R}(\tau_{R})$ from the usual string theory. The following quantity (\ref{8.1}) is introduced for each ``object'' taken to cover/correspond to a $25+1$-vector (\ref{diffBl55}),
\begin{eqnarray}
J^{\mu}(I)=X^{\mu}_{R}\left( \tau_{R}\left(I+\frac{1}{2}\right) \right)-X^{\mu}_{R}\left( \tau_{R}\left( I-\frac{1}{2}\right) \right)
\label{8.1}
\end{eqnarray}
by taking a small interval $\triangle\tau_{R}$ (corresponding to object number $I$) in the $\tau_{R}=\tau-\sigma$ variable as interval for which the $\tau_{R}$-value $\tau_{R}\left(I+\frac{1}{2}\right)$ is the upper end and $\tau_{R}\left(I-\frac{1}{2}\right)$ is the lower end. For a theory with both open and closed strings we have only one type of ``objects'' and we use them for both right- and left- movers so that we also have objects given as
\begin{eqnarray}
J^{\mu}(I)=X^{\mu}_{L}\left(\tau_{L}\left(I+\frac{1}{2}\right)\right)-X^{\mu}_{L}\left(\tau_{L}\left(I-\frac{1}{2}\right)\right).
\end{eqnarray}
But for theories with only closed strings we shall distinguish two types of ``objects'', right-mover ones $J^{\mu}_{R}(I)$ and left-mover ones $J^{\mu}_{L}(I)$.

In addition to the classification into ``cyclically ordered chains'' of ``objects'' each chain corresponding to an open string, we divide the ``objects'' sitting along these alternating to be even, odd, even, odd, even,c. An important ``technical detail'' is that we only consider the even objects as truly existing and construct our Fock space (meaning the Hilbert space of the second quantized string theory = string field theory) alone from the \underline{even} ``objects''. The odd objects do not count as truly existing and are rather just made as mathematical constructions from the nearest neighboring (on the chains) conjugate variables $\Pi^{i}(I)$ of the even ``objects'' $J^{i}(I)$. Since the concept of ``nearest neighbors'' giving rise to the odd objects from even ones is derived from the ``cyclically ordered chains'' and thus (corresponding to the open strings) the ``odd objects'' to be constructed in a way, the ``odd objects'' are in themselves not considered ontologica
 lly existing in our model. 

The fundamental degrees of freedom of the even ``objects'' are only the $24$ ``transverse degrees of freedom'' and to match $25+1$ degrees of freedom, two extra dimensions are constructed: The +-component in the infinite momentum frame formation is taken to be proportional to the number of ``objects'' (when the number of ``objects'' is concerned it does not matter if we count only even or both odd and even because there are always equally many even and odd ``objects'').

The $-$component is adjusted or rather defined so that all the $\bigl( J^{\mu}(I) \bigr)^{2}=0$, i.e. all the ``objects'' are light-like vectors. This requirement actually corresponds to the constraints in string theory.

In theories with closed strings - for which we did not go deeply - one has to have two cyclically ordered chains for each closed string, one right-mover one and one left-mover one. If we have a theory with only closed strings these two cyclically ordered chains have to consist of two different types of objects. So we need in that case two completely different types of ``objects''. Only if the theory has also open strings these two classes of ``objects'' are united to only one type.

Our main results were that strings represented by ``objects'' of our type obtained a mass square operator which look to mean exactly the spectrum of the open string in the ``continuum'' limit which indicates that the number of ``objects'' in a cyclically ordered chain $N$ is much larger than the vibration mode number $L$, i.e. $N>>L$.

There were though an interesting technical difficulty in our calculation of this spectrum: At first we seemingly rather than $24$ oscillation modes for each frequency $\omega=\frac{L}{\alpha'}$ we obtained $2 \cdot 24$. We argued that this difficulty should be solved by requiring for the to be taken as ``smooth'' modes that there should be smoothness not only of the variables $J^{\mu}(I)$ and $\Pi^{\mu}(I)$ restricted to even ``objects'', but also required smoothness extended to odd as well as even ``objects''. Really we could look on the seemingly too many degrees of freedom. popping up compared to string theory at first as a species doubler problem due to our discretization.

So far thus our proposal for our string field theory looks extremely promising by delivering the right spectrum for string theory.

We have further already put forward the prescription for how to construct scattering amplitudes - using in fact a philosophy that ``nothing happens under the scattering'' - by prescribing that one shall just write down the wave function of the incoming and outgoing states in terms of our even ``objects''. Then the scattering amplitude or $S$-matrix is expected to simply appear as the overlap between these two states. We have not yet completely performed this Veneziano model calculation \cite{21}.

There are also some ``small'' difficulties to be settled to finish that calculation completely. Especially we have some intriguing problems to obtain conservation of the $-$component $P^{-}$ of the ($25+1$)-momentum, i.e. the energy in the infinite momentum frame. But generally there are indications that the Veneziano model will indeed result. 

When these ``minor'' difficulties are overcome we shall have shown that our string field theory model is indeed identical to string theory with an arbitrary number of strings. That would then mean that our model can be considered at the roof of string theory. But our model is at what we call the ``fundamental level'', i.e. only the transverse $24$ components of the even ``objects'', exceedingly trivial. So our scheme may as well as a string field theory be considered a solution of the then solvable string theory. This solvability might well be behind many of the great achievements of string theories, such as Maldacena conjecture \cite{23}.

\begin{acknowledgments}
The authors acknowledge K. Murakami for his useful comments and informing them references on string field theory.
One of us (H. B. N.) thanks the Niels Bohr Institute for allowance to work as emeritus and for Masao to visit to give a talk in connection with the ``Holger Fest''. Masao Ninomiya acknowledges the Niels Bohr Institute and the Niels Bohr International Academy for inviting to the ``Holger Fest''. He also acknowledges that the present research is supported in part by the J.S.P.S. Grant-in-Aid for Scientific Research Nos.21540290, 23540332 and 24540293. Also H. B. N. thanks to the Bled conference participants organizers and Matiaz Breskov for financial  support to come there where many of the ideas of this work got tested again.
\end{acknowledgments}


\end{document}